\documentclass[11pt]{article}
\newcommand{\newsection}[1]{\section{#1}\setcounter{equation}{0}}

\newcounter{newapp}
\usepackage{amssymb}
\usepackage{euscript}
\def\l{\lambda}
\def\a{{\alpha}}
\def\b{{\beta}}

\def\l{\lambda}

\def\k{\kappa}
\def\t{{\mathbf t}}

\def\v{{\mathbf v}}
\def\e{{\epsilon}}
\def\bra#1{\langle #1 |}
\def\ket#1{|#1 \rangle}
\def\0{\nonumber}

\def\det{{\rm det}}
\def\Det{{\rm Det}}
\def\log{{\rm log}}

\def\exp{{\rm exp}}

\newcommand\T{{\cal{T}}}
\newcommand\Q{{\cal{Q}}}
\newcommand\K{{\cal{K}}}
\newcommand\EA{\EuScript{A}}
\newcommand\EB{\EuScript{B}}
\newcommand\EC{\EuScript{C}}
\newcommand\ED{\EuScript{D}}
\newcommand\EF{\EuScript{F}}
\newcommand\EG{\EuScript{G}}
\newcommand\EH{\EuScript{H}}

\newcommand\EP{\EuScript{P}}

\newcommand\V{{\cal{V}}}

\newcommand\M{{\cal{M}}}
\newcommand\N{{\cal{N}}}
\newcommand\CS{{\cal{S}}}
\newcommand\CP{{\cal{P}}}

\newcommand\be{\begin{eqnarray}}
\newcommand\ee{\end{eqnarray}}            
\newcommand\ba{\begin{array}}           
\newcommand\ea{\end{array}}
\newcommand\beq{\begin{equation}}
\newcommand\eeq{\end{equation}}           

\begin{document}
\begin{flushright}
 SISSA/28/04/EP\\ ZTF--04--01
\end{flushright}

\vspace{0.9in}
\begin{center}
{\Large\bf The Perturbative Spectrum of the Dressed Sliver}
\end{center}
\vspace{0.1in}
\begin{center}
{\bf L. Bonora}\footnote{ bonora@sissa.it},
{\bf C. Maccaferri}\footnote{maccafer@sissa.it},
\vspace{1mm}

{\it International School for Advanced Studies (SISSA/ISAS)\\
Via Beirut 2--4, 34014 Trieste, Italy,}\\ and\\{\it INFN, Sezione di
Trieste}\\
\vspace{2mm}
{\bf P. Prester}\footnote{pprester@phy.hr}\\
\vspace{1mm}
{\it Max-Planck-Institut f\"{u}r Gravitationsphysik
(Albert-Einstein-Institut)\\
Am M\"{u}hlenberg 1, D-14476 Golm (b.\ Potsdam), Germany}\\ and \\
{\it Theoretical Physics Department, Faculty of Science,
University of Zagreb\\
Bijenicka c.\ 32, p.p.\ 331, HR-10002 Zagreb, Croatia}
\end{center}
\vspace{0.3in}
\begin{center}
{\bf Abstract}
\end{center}
We analyze the fluctuations of the dressed sliver solution found
in a previous paper, hep-th/0311198, in the operator formulation
of Vacuum String Field Theory. We derive the tachyon wave function
and  then analyze the higher level fluctuations. We show that the
dressing is responsible for implementing the transversality
condition on the massless vector. In order to consistently deal
with the singular $k=0$ mode we introduce a string midpoint
regulator and we show that it is possible to accommodate all the
open string states among the solutions to the linearized equations
of motion. We finally show how the dressing can give rise to the
correct ratio between the energy density of the dressed sliver and
the brane tension computed via the three--tachyons--coupling.
~\\
\vspace{5cm} \pagebreak

\newsection{Introduction}

In a companion paper, \cite{BMP1}, in a search for nonperturbative
solutions with the appropriate features to represent D-branes, we
found a solution of Vacuum String Field Theory (VSFT), the dressed sliver,
endowed with some interesting properties, the most important being the
possibility to define, via a suitable regularization, a finite energy density.
The purpose of this paper is to analyze the
perturbative spectrum around this solution.

Before we plunge into the technicalities of the analysis of the
spectrum, it is worth summarizing the motivations for this research.
VSFT, \cite{Ras, RSZ1}, is conjectured to represent SFT, \cite{W1}, at
the tachyon condensation vacuum, and it is expected to represent a new
theory, radically different from the SFT constructed on the initial
(unstable) vacuum. It was argued that this new vacuum can only be the
critical bosonic closed string vacuum. If so it should harbor
nonperturbative solutions representing (unstable) D-branes of any
dimensions. As it turned out, such solutions were found in copious
variety, \cite{KP, RSZ2, RSZ3, GRSZ2, HKw}, lending evidence to the
above conjecture. A particularly simple and appealing one (both from the
algebraic and geometric point of view)  was the sliver.

However some important problems have remained unsolved. First, the
energy density
of these solutions, in particular of the sliver, is evanescent
and one is obliged to postulate an infinite multiplicative
constant in front of the action to account for a physical solution,
\cite{Oku2, GRSZ1, Sch1, Oka2}.
Second, the attempts at finding the perturbative spectrum
around these solutions (i.e. the spectrum of the open strings
attached to the D-branes) have been successful in the BCFT approach,
\cite{RSZ4, Oka}, but it has proven to be remarkably hard (particularly
in regards to the task of reconstructing the Virasoro constraints)
in the operator approach, \cite{HM3, HK}.

These failures of the operator approach should not be underestimated.
The issue here is not the
practical use one can make of the spectrum of, say, the D25-brane
obtained in this way, i.e. of the fluctuations around
a nonperturbative solution of VSFT. If one wishes to describe
some physics of the D25-brane, one had better use the formalism
of the initial SFT about the trivial vacuum. The crucial issue here
is rather the consistency of VSFT itself. Let us recall that
VSFT is a simplified version of SFT, the simplification being
determined by the very simple form taken by the BRST charge
$\Q$, which is expressed only in terms of the ghost oscillators
$c_n$ (see below). This may at first look as an oversimplification:
thanks to it one can derive in a simple and elegant manner
the abovementioned nonperturbative solutions, but one may have
to pay a cost in terms of loss of information, which seems to
show up when we come to determining the spectrum. To sum up:
can we regard the operator formulation of VSFT as a reliable approach or not?
We would like to explore in this paper the possibility that the
answer to this question is yes. We believe in fact that the true
reason for the above failures is that the sliver solution is
too singular and need to be regularized, but once this is properly done
the above-mentioned problems disappear.

We have proven in \cite{BMP1} that, using a new type of solution,
akin to the sliver, but somehow deformed with respect to it,
which we dubbed dressed sliver, the problem of the energy density
may find a solution. In this
paper we continue the analysis started there, by studying the
perturbative spectrum around the dressed sliver, i.e. the solutions
to the linearized equations of motion (LEOM). We prove
that not only can the full open string mass spectrum
be recovered, but that the right Virasoro constraints
come out naturally from the LEOM. The key ingredients that make this
possible are, on the one hand, the dressing
and, on the other hand, a careful analysis about the string midpoint.
These are the new elements we introduce in the analysis with respect to
\cite{HKw}, which otherwise we follow rather closely.
In this paper we limit ourselves to the D25--brane case, although,
as shown in \cite{BMP1},
the analysis can be extended to lower dimensional branes.

This result seems to justify on a more substantial basis the
claim that VSFT has solutions that can be interpreted as
D-branes of any dimensions. The existence of such solutions are
expected on the basis of the lore that D-branes are sources
of closed strings and on the conjecture that the tachyon
condensation vacuum corresponds to the closed string vacuum.
In turn this add support to our philosophy that VSFT is
a complete theory. As an aside, we notice that
the D-brane solutions being expressed in terms
of the (initial) open string creation operators, may be considered
a clear manifestation of the open-closed string duality
advocated by A. Sen, \cite{Senroll}.

It is fair now to also mention that we
have not been able to find an algorithmic way to construct all the
solutions to the linearized equations of motion. As a consequence we
have explicitly analyzed only the first few levels, although there
seems to be no conceivable obstacles to the extension to higher
levels. Our analysis of the cohomological structure, still far from
being exhaustive, confirms that for dressed sliver solutions,
physical excitations are concentrated around the `midpoint' $k=0$ of
the continuous basis, as claimed in \cite{ima, HK} for
previous classical solutions of VSFT. We also find an
additional set of solutions: they are constructed by means
of polynomials of $\xi a^\dagger$  ($\xi$ is the dressing
vector) applied to any physical state. As a consequence any physical
state is accompanied by an infinite tower of descendants
which seem to contain the same information as far as observables
are concerned. These extra solutions might be remnants of
the full gauge symmetry of VSFT.

The paper is organized as follows. Section 2 is a review about the
dressed sliver solution. Section 3 is a
general and preliminary presentation of the LEOM  in VSFT: we
introduce there all the general objects which are needed in the
course of our analysis. In section 4 we find solutions for the
infinite vector $\t$ and the number $G$ which play a crucial role
in the construction of all the open strings states as solutions to
the LEOM.

Next, in section 5, we analyze the lowest level excitations: the
tachyon state is found at the correct mass shell. The vector
excitation is also rather straightforward: we show not only how to
define the corresponding string field solution but, in particular, how
transversality is implemented thanks to dressing. In order to continue
with the analysis of higher level string modes it turns out that the
relevant information is concentrated around the midpoint $k=0$ of the
continuous basis. We therefore need a technique to probe this region.
To this end in section 6 we introduce a new regulator $\eta$
by means of which we are able to control the singularity at the $k$-basis
midpoint. In section 7 we determine the solutions corresponding
to the string modes of level 2 and 3 of the canonical quantization.
In section 8 we briefly discuss the cohomology properties of the
previous states. In section 9 we show that states that represent
physical modes are accompanied by an infinite tower of dressing
excitations which, as far as we could verify, does not alter any physical
observable. In section 10, we evaluate the three-tachyon coupling and
comment on how the well-known puzzle that links such coupling to the
energy density of the classical solution may be resolved in our
approach, thanks to the dressing. The copious calculations that
underpin the above results are presented or summarized in several
appendices.

\newsection{The dressed sliver: a review}

In order to render this paper as self--contained as possible,
in this section we review the main properties of the dressed sliver
solution.

To start with we recall some formulas relevant to VSFT. The action is
\beq
{\cal S}(\Psi)= - \frac 1{g_0^2} \left(\frac 12 \langle\Psi |
{\cal Q}|\Psi\rangle +
\frac 13 \langle\Psi |\Psi *\Psi\rangle\right)\label{sftaction}
\eeq
where
\beq
{\cal {Q}} =  c_0 + \sum_{n>0} \,(-1)^n \,(c_{2n}+ c_{-2n})\label{calQ}
\eeq
The equation of motion is
\beq
{\cal Q} \Psi = - \Psi * \Psi\label{EOM}
\eeq
and the ansatz for nonperturbative solutions has the factorized form
\beq
\Psi= \Psi_m \otimes \Psi_g\label{ans}
\eeq
where $\Psi_g$ and $\Psi_m$ depend purely on ghost and matter
degrees of freedom, respectively. Then eq.(\ref{EOM}) splits into
\be
{\cal Q} \Psi_g & = & - \Psi_g *_g \Psi_g\label{EOMg}\\
\Psi_m & = & \Psi_m *_m \Psi_m\label{EOMm}
\ee
where $*_g$ and $*_m$ refers to the star product involving only the ghost
and matter part.\\
The action for this type of solution becomes
\beq
{\cal S}(\Psi)= - \frac 1{6 g_0^2} \langle \Psi_g |{\cal Q}|\Psi_g\rangle
\langle \Psi_m |\Psi_m\rangle \label{actionsliver}
\eeq
$\langle \Psi_m |\Psi_m\rangle$ is the ordinary inner product,
$\langle \Psi_m |$ being the $bpz$ conjugate of $|\Psi_m\rangle$
(see below).

 The $*_m$ product is defined as follows
\beq
_{123}\!\langle V_3|\Psi_1\rangle_1 |\Psi_2\rangle_2 =_3\!
\langle \Psi_1*_m\Psi_2|
\label{starm}
\eeq
where the three strings vertex $V_3$ is
\beq
|V_3\rangle_{123}= \int d^{26}p_{(1)}d^{26}p_{(2)}d^{26}p_{(3)}
\delta^{26}(p_{(1)}+p_{(2)}+p_{(3)})\,{\rm exp}(-E)\,
|0,p\rangle_{123}\label{V3}
\eeq
with
\beq
E= \sum_{a,b=1}^3\left(\frac 12 \sum_{m,n\geq 1}\eta_{\mu\nu}
a_m^{(a)\mu\dagger}V_{mn}^{ab}
a_n^{(b)\nu\dagger} + \sum_{n\geq 1}\eta_{\mu\nu}p_{(a)}^{\mu}
V_{0n}^{ab}
a_n^{(b)\nu\dagger} +\frac 12 \eta_{\mu\nu}p_{(a)}^{\mu}V_{00}^{ab}
p_{(b)}^\nu\right) \label{E}
\eeq
Summation over the Lorentz indices $\mu,\nu=0,\ldots,25$
is understood and $\eta$ denotes the flat Lorentz metric.
The operators $ a_m^{(a)\mu},a_m^{(a)\mu\dagger}$ denote the non--zero
modes matter oscillators of the $a$--th string, which satisfy
\beq
[a_m^{(a)\mu},a_n^{(b)\nu\dagger}]=
\eta^{\mu\nu}\delta_{mn}\delta^{ab},
\quad\quad m,n\geq 1 \label{CCR}
\eeq
$p_{(r)}$ is the momentum of the $a$--th string and
$|0,p\rangle_{123}\equiv |p_{(1)}\rangle\otimes
|p_{(2)}\rangle\otimes |p_{(3)}\rangle$ is
the tensor product of the Fock vacuum
states relative to the three strings with definite c.m.  momentum .
$|p_{(a)}\rangle$ is annihilated by the annihilation
operators $a_m^{(a)\mu}$ ($m\geq1$) and it is eigenstate of the
momentum operator $\hat p_{(a)}^\mu$
with eigenvalue $p_{(a)}^\mu$. The normalization is
\beq
\langle p_{(a)}|\, p'_{(b)}\rangle = \delta_{ab}\delta^{26}(p+p')\0
\eeq

The symbols $V_{nm}^{ab}$  denote
the coefficients computed in \cite{GJ1, GJ2, Ohta,
tope, leclair1, leclair2}.
We will use them in the notation of Appendix A and B of \cite{RSZ2}.

To complete the definition of the $*_m$ product we must specify the
$bpz$ conjugation properties of the oscillators
\beq
bpz(a_n^{(a)\mu}) = (-1)^{n+1} a_{-n}^{(a)\mu}\0
\eeq

The sliver solution to eq.(\ref{EOMm}) is given by
\beq
|\Xi\rangle = \N e^{-\frac 12 a^\dagger Sa^\dagger}\ket{0},\quad\quad
a^\dagger S a^\dagger = \sum_{n,m=1}^\infty a_n^{\mu\dagger} S_{nm}
 a_m^{\nu\dagger}\eta_{\mu\nu}\label{Xi}
\eeq
where the Neumann matrix $S$ is as follows. Let us introduce the twisted
matrices $X=CV^{11},X_+=CV^{12}$
and $X_-=CV^{21}$, together with $T=CS=SC$, where
$C_{nm}= (-1)^n\delta_{nm}$. They are mutually commuting and $T$
is given by
\beq
T= \frac 1{2X} (1+X-\sqrt{(1+3X)(1-X)})\label{sliver}
\eeq

The normalization constant $\N$ is
\beq
\N= (\Det (1-\Sigma \V))^{\frac{D}{2}}\label{norm}
\eeq
and
\beq
\Sigma= \left(\matrix{S&0\cr 0& S}\right),
\quad\quad\quad
{\cal V} = \left(\matrix{V^{11}&V^{12}\cr V^{21}&V^{22}}\right),
\label{SigmaV}
\eeq
The norm of the sliver is
\beq
\langle \Xi|\Xi\rangle = \frac {\N^2}{(\det (1-S^2))^{\frac{D}{2}}}
\label{ener}
\eeq
Both eq.\ (\ref{norm}) and (\ref{ener}) are ill--defined and need to
be regularized, after which they both result to be vanishing,
\cite{Oku2}.

The dressed sliver is a sort of deformation of the sliver, which is
obtained as follows. We first introduce the infinite real vector
$\xi=\{\xi_{n}\}$
which is  chosen to satisfy the condition
\beq
\rho_1 \xi =0,\quad\quad \rho_2 \xi =\xi, \label{xi1}
\eeq
where $\rho_1(\rho_2)$ are the Fock space projectors into the right
(left) half of the string (see Appendix A).
Next we set
\beq
\langle\xi| \frac 1{1-T^2}|\xi\rangle =1 ,\quad\quad
\langle\xi| \frac {T}{1-T^2}|\xi\rangle= \k
\label{noncond}
\eeq
The first equation sets the normalization of $\xi$.
The second tells us that $\k$ is a real negative number, see \cite{BMP1}.

The {\it dressed sliver} solution is given by an
ansatz similar to (\ref{Xi})
\beq
|\hat\Xi\rangle = \hat{\cal N} e^{-\frac 12 a^\dagger \hat S a^\dagger}\ket 0,
\label{Xihat}
\eeq
with $S$ replaced by
\beq
\hat S = S +R,\quad\quad R_{nm}= \frac 1{\kappa +1}\left(\xi_n(-1)^m\xi_m
+\xi_m(-1)^n\xi_n\right)\label{Shat}
\eeq
Alternatively
\beq
\hat T = T +P,\quad\quad  P=\frac 1{\kappa +1}
\left(|\xi\rangle\langle\xi|+|C\xi\rangle\langle C\xi|
\right) \label{That}
\eeq
As shown in \cite{BMP1} the matrix $\hat S$ satisfies the equation
\beq
\hat S=V^{11} +(V^{12},V^{21})({1}-
\hat\Sigma{\cal V})^{-1}\hat\Sigma
\left(\matrix{V^{21}\cr V^{12}}\right)\label{hatShatS}
\eeq
where
\beq
\hat\Sigma= \left(\matrix{\hat S&0\cr 0& \hat S}\right),\label{hatSigma}
\eeq
which is required for $\hat \Xi$ to be a projector.

The normalization is given by
\beq
\hat \N= [\Det(1- \Sigma \V)]^{D/2}
\left(\frac {1}{\k+1}\right)^{\!D}\label{hatN}
\eeq
This determinant, after regularization, turns out to vanish, \cite{Oku2}. The action
corresponding to $\hat \Xi$ is ill--defined.
In \cite{BMP1} we devised a way to regularize the action by introducing
a deformation parameter $\e$ multiplying $R$ as follows
\beq
|\hat\Xi\rangle \to |\hat\Xi_\e\rangle,\quad\quad
\hat S \to \hat S_\e = S +\e R\label{drsl}
\eeq
Therefore $\hat \Xi_\e$ coincide with the sliver for $\e=0$ and with
the dressed sliver for $\e=1$. For $\e \neq0,1$ the new state is in
general not a projector. However its limit for $\e\to 1$ allows us to
define a finite norm  $\langle \hat \Xi|\hat\Xi\rangle$.
This goes according to the following recipe.
We regularize the determinants involving the sliver matrices $X, X_+,X_-$
and $T$ by truncating them at a finite level $L$. In the limit
$L\to\infty$ these determinants behave like fractional powers of $L$.
The next ingredient consists in using the nested  limits prescription
\beq
\lim_{\e_1\to 1}\left(\lim_{\e_2\to 1}
\langle \hat \Xi_{\e_1} |\hat \Xi_{\e_2} \rangle\right)\label{limlim}
\eeq
which gives
\be
&&\frac 1{\langle 0|0\rangle}\,\lim_{\e_1\to 1}\left(\lim_{\e_2\to 1}
\langle \hat \Xi_{\e_1} |\hat \Xi_{\e_2}\rangle\right) \label{xi1xi2}\\
&&= \lim_{\e_1\to 1}
\left(\frac{1}{(\k+1)^2 }\right)^{\!\frac D2}
\left(\frac {L^{-\frac 5{36}}}{1-\e_1}\right)^{\!D}+\ldots =
\left(\frac{1}{(\k+1)^2 s^2}\right)^{\!\frac D2}\label{mattlim}
\ee
provided we use the last ingredient, i.e. the tuning
\beq
1-\e_1 =s L^{-\frac 5{36}}\label{regpres1}
\eeq
in the limit $\e_1\to 1,L\to\infty$. This recipe guarantees that
the equation of motion is satisfied when evaluated on a very
large class of states (including the entire Fock space and the dressed
sliver itself).

The ghost part of the solution, i.e. a solution to (\ref{EOMg}),
can be found much in the same way. It has the form of a squeezed state
\begin{equation} \label{ghdsl}
|\widehat{\widetilde{\Xi}}_{\tilde{\epsilon}}\rangle
= \widehat{\widetilde{\N}}_{\tilde{\epsilon}} \,
e^{c^\dagger \widehat{\widetilde{S}}_{\tilde{\epsilon}} b^\dagger}
c_1 |0\rangle
\end{equation}
with
\be
\widehat{\widetilde{S}}_{\tilde{\epsilon}} =
\widetilde{S} + \tilde{\epsilon}\widetilde{R} \;,
\qquad
\widetilde{R} = \frac{1}{\widetilde{\kappa}+1}
 (|C\delta\rangle \langle\beta| + |\delta\rangle \langle C\beta|)\0
\ee
for $\e=1$. Here we have introduced directly the deformation parameter
$\tilde\e$, which is necessary in order to regularize the action.
The vectors $\beta$ and $\delta$ are such that
\begin{equation}
\widetilde{\rho}_1 \beta
= \widetilde{\rho}_1 \delta = 0 ,\quad\quad \widetilde{\rho}_2 \beta=\beta,
\quad\quad  \widetilde{\rho}_2 \delta = \delta \;.\label{rhotilde}
\end{equation}
with
\begin{equation}
\langle\beta| \frac{1}{1-\widetilde{T}^2} |\delta\rangle = 1 \;,
\qquad
\langle\beta| \frac{\widetilde{T}}{1-\widetilde{T}^2} |\delta\rangle
= \widetilde{\kappa}
\end{equation}
The first equation fixes the normalization of $\beta$ and $\delta$,
and the second is the definition of $\widetilde{\kappa}$.
Once again, the quantity
$\langle \widehat{\widetilde{\Xi}}_{1}|
\widehat{\widetilde{\Xi}}_{1}\rangle$ is ill--defined but, by using an
ordered limiting prescription, one can show that
\begin{equation}
\lim_{\tilde{\epsilon}_1\to1} \left( \lim_{\tilde{\epsilon}_2\to1}
\langle\widehat{\widetilde{\Xi}}_{\tilde{\epsilon}_1}|
\mathcal{Q} \, |\widehat{\widetilde{\Xi}}_{\tilde{\epsilon}_2}\rangle \right)
= (\widetilde{\kappa}+1)^2 \, \tilde{s}^2\;.\label{ghostlim}
\end{equation}
provided
\begin{equation}
1-\tilde{\epsilon}_1 = \tilde{s} L^{-\frac{11}{36}}
\end{equation}
for some constant $\tilde{s}$.
Eq.\ (\ref{ghostlim}) together with (\ref{mattlim}) allow us to define
a finite value for the action (\ref{actionsliver}). This
relates the physical coupling constant to the regularization
procedure, see below.

From now on, to simplify the notation, we often understand
the limit for $\e\to 1$. In the following we will therefore
represent the dressed sliver solution as the factorized product
$|\hat \Xi\rangle \otimes |\widehat{\widetilde\Xi}\rangle$.
We will however reintroduce the $\e$--dependence whenever a risk of
ambiguity arises.

To end this summary we would like to make a comment on the eigenvalues
of the Neumann matrix of the dressed sliver, which hopefully
clarifies some of the formulas used below. As we have remarked,
this Neumann matrix does not commute with the sliver matrix $T$, so they
cannot share their eigenvectors. However much can be said about
the eigenvalues of $\hat T$. If the vector $\xi$ is square--summable
(as we suppose), $P$ is a compact operator. Perturbing $T$ by a
compact operator does not modify its continuous spectrum, \cite{fa}.
Therefore $\hat T$ must have the same continuous spectrum as $T$.
In addition, however, it might have isolated eigenvalues of its own.
It is easy to show that $\hat T$ does develop an extra discrete eigenvalue 1.
This fact can be easily guessed from the result of \cite{BMP1}
\be
\det(1-\hat{T_\e})=(1-\e)^2\det(1-T),
\ee
which suggests that $\hat T$ has a doubly degenerate eigenvalue 1.
It turns out that the corresponding eigenvectors have definite twist
and are given by
\beq
\ket{\chi_\pm}=\frac{1}{1-T}(1\pm C)\ket\xi\label{extra}
\eeq
as can be easily proved by applying (\ref{That}) to the above expression.

This is in fact the reason why the $bpz$ norm of the dressed sliver can be made
finite by appropriately tuning the vanishing behavior induced by the midpoint
$k=0$ and the divergent one induced by this discrete eigenvalue.
We will see, in the study of the spectrum, that these new eigenvectors are
responsible for creating  an infinite tower
of ``descendants''of every physical state, with same mass and same
polarization conditions as the initial state.

\newsection{The linearized equation of motion}

Let us call for simplicity
$\Phi_0=|\hat\Xi\rangle\otimes|\widehat{\widetilde\Xi}\rangle$
the overall (matter+ghost)
solution we have just reviewed. If we write $\Psi=\Phi_0+\phi$,
the action becomes
\beq
{\cal S}(\Psi)= {\cal S}(\Phi_0)- \frac 1{g_0^2}
\left(\frac 12 \langle\phi |{\cal Q}_0|\phi\rangle +
\frac 13 \langle\phi |\phi *\phi\rangle\right)\label{sftaction'}
\eeq
where
\beq
{\cal {Q}}_0 \phi= \Q\phi +\Phi_0*\phi+\phi*\Phi_0 \label{calQ0}
\eeq
The equation of motion for small fluctuations around the solution
$\Phi_0$ is therefore
\beq
\Q\phi +\Phi_0*\phi+\phi*\Phi_0 =0\label{EOMl}
\eeq
The solutions to this linearized equation of motion (LEOM)
are expected to encompass all the modes
of the open strings with endpoints on the D25--brane represented by
$\Phi_0$ as well as all the states which are $\Q_0$--exact.

To find the solutions to (\ref{EOMl}) we follow \cite{HKw}, but we
introduce some significant changes: the dressing and the midpoint
regularization.
The ansatz for a general solution of momentum $p$ is as follows
\beq
|\hat\phi_{e}(\EP,\t,p)\rangle = \N_e \EP(a^\dagger)\,
\exp{[-\sum_{n\geq 1}
t_n a_n^{\mu\dagger}\hat p_\mu}]
|\hat \Xi_e\rangle \otimes
|\widehat{\widetilde\Xi}\rangle{\rm e}^{ipx}\equiv
|\varphi_e(\EP,\t,p)\rangle\otimes
|\widehat{\widetilde\Xi}\rangle\label{ansatz}
\eeq
where $\t=\{t_n\}$, $\EP(a^\dagger)$ is some polynomial of expressions
of the type $\sum_n \zeta_n a^\dagger_n$, and
\be
\hat p \,{\rm e}^{ipx} = p\,{\rm e}^{ipx},\quad\quad bpz(\hat p)= -\hat p\0
\ee

We will often drop the labels $\t,\EP$ and $p$ when no ambiguities are
possible. The factorized form of $(\ref{ansatz})$ allows us to split the linearized
equation of motion into ghost and matter part
\be
&&\Q |\widehat{\widetilde\Xi}\rangle + |\widehat{\widetilde\Xi}\rangle
*_g|\widehat{\widetilde\Xi}\rangle=0\label{EOMlg}\\
&&|\hat\varphi_e\rangle = |\hat \Xi\rangle *_m |\hat\varphi_e\rangle +
|\hat\varphi_e\rangle *_m |\hat \Xi\rangle \label{EOMlm}
\ee
The ghost part will remain the same throughout the paper, and from now
on we simply forget it and concentrate on the matter part.

In the above equation $|\hat \Xi_e\rangle$ formally coincides with
$|\hat \Xi_\e\rangle$,
with $\e$ replaced by $e$. The reason for this seemingly bizarre change
of notation is because  the parameter $e$ plays a different role
from $\e$. While $\e$ is a deformation parameter and we are only interested
in the limit $\e\to 1$ (recall that for $\e\neq 0,1\,$ $\hat\Xi_\e$ is not a
solution to (\ref{EOMm})), we will find that the linearized equation of
motion can be solved for any value of $e$. The reason of this
lies in a result we found in \cite{BMP1}, see eq.(4.15)
there,
\beq
\ket{\hat\Xi_\e}*\ket{\hat\Xi_e}=
\ket{\hat\Xi_{\e\star e}}\label{nostruc}
\eeq
where $\ket{\hat\Xi_\e}$ is the same as in (\ref{drsl}) with
\be
\hat\N_\e = \N\left(\frac{1+(1-\e)\k}{\k+1}\right)^{\!D},\quad\quad
\hat\N_e = \N\left(\frac{1+(1-e)\k}{\k+1}\right)^{\!D}\0
\ee
and
\beq
\e \star e = \frac {\e e}{1+(1-\e)(1-e)\k}\label{estareta}
\eeq
The $\star$-multiplication is isomorphic to ordinary multiplication between
real numbers: using the reparametrization
\begin{equation}
f_\epsilon = \frac{1+(1-\epsilon)\kappa}{\epsilon}
= 1 + (\kappa+1)\frac{1-\epsilon}{\epsilon}
\end{equation}
it is easy to check that $f_{\epsilon \star e}=f_\epsilon f_e$.

It is evident from (\ref{nostruc}) that
\beq
\ket{\hat\Xi_e}*\ket{\hat\Xi}=\ket{\hat\Xi}*\ket{\hat\Xi_e}=
\ket{\hat\Xi_{e}}\label{nostruc'}
\eeq
for any value of the parameter $e$. This basic equality will allow us to
construct solutions to the LEOM that contain the free parameter $e$.
We anticipate that eventually, in order to guarantee finiteness of the
three-tachyons coupling, $e$ will have to be set to 1.

Let us see all this in more detail, i.e.
let us find the general conditions for solving the LEOM
(\ref{EOMl}). To this end we introduce the general state
\beq
|\hat\varphi_{e,\beta}\rangle \,=\,\N_e \, \exp{\Big[\! -\sum_{n\geq 1}
t_n a_n^{\mu\dagger}\hat p_\mu}
- \sum_{n\geq 1} \beta^{\mu}_n a_n^{\nu\dagger} \eta_{\mu\nu} \Big]
|\hat \Xi_e\rangle\, {\rm e}^{ipx}
\label{genans}
\eeq
where, with respect to (\ref{ansatz}), we have inserted the
parameters $\beta^{\mu}_n$. By differentiating with respect to it the
appropriate number of times and setting afterwards $\beta^\mu_n=0$, we will
be able to generate any polynomial in $a_n^{\dagger}$ and therefore
reproduce any state of the form (\ref{ansatz}).

Now we need
\be
&&_1\langle \hat\Xi_\e|_2\langle \hat \varphi_{e,\beta} |V_3\rangle=
\frac{\hat\N_\e \hat\N_e}{(\det \hat\K_{\e e})^{\frac D2}}
\exp{\left[-\chi^T \hat\K_{\e e}^{-1}\lambda -
\frac 12 \chi^T \hat\K_{\e e}^{-1}\chi
-\frac 12 \lambda^T \V \hat\K_{\e e}^{-1} \lambda \right]}\0\\
&&~~~~~~~\cdot \,\exp{\left[-\frac 12 \sum_{n,m\geq 1} a_n^{(3)\dagger}
V_{n,m}^{33} a_m^{(3)\dagger} - a_n^{(3)\dagger}
(\v_{0n}-\v_{+n})p\right]}
|0\rangle_3 {\rm e}^{-p V_{00}p}\, {\rm e}^{ipx}\label{genstar1}
\ee
where we introduced
\be
\hat\K_{\e e} = 1- \hat\CS_{\e e} \V,\quad\quad \hat\CS_{\e e} =
\left(\matrix{\hat S_\e &0\cr 0&\hat S_e\cr}\right)\0
\ee
together with
\beq
\chi= \left(\matrix{V^{21} a^{(3)\dagger} + p (\v_+ -\v_-)\cr
V^{12} a^{(3)\dagger} + p (\v_{-} -\v_0)\cr}\right), \quad\quad
\lambda = C\left(\matrix{0\cr {\mathbf\beta} - p\t}\right)
\label{chilambda}
\eeq
In all these formulas we have introduced infinite vectors
${\mathbf\beta^\mu},\,\t,\,\v_0,\,\v_+,\,\v_-$ with components
\beq
\beta_n^\mu,\quad\quad t_n,\quad\quad
\v_{0n} = V^{11}_{0n}= V^{22}_{0n}, \quad\quad
\v_{+n} = V^{12}_{0n},\quad\quad \v_{-n}= V^{21}_{0n},\label{vn}
\eeq
respectively.
We are interested in the above formula in the limit $\e\to 1$, while
keeping $e$ fixed.

Let us recall from \cite{BMP1}, Appendix B.2, that
\be
&&\hat \N_\e = [\Det(1- \Sigma \V)]^{D/2}
\left(\frac {f_\e}{\k+f_\e}\right)^{\!D}\0\\
&&\Det(1-\hat\Sigma_{\e e}\V)=\left(\frac{\k+f_\e f_e}{(\k+f_\e)(\k+f_e)}
\right)^2\Det(1-\Sigma \V)\0
\ee
from which we get the important relation
\beq
\lim_{\e\to 1} \frac{\hat\N_\e}{(\sqrt{\det \hat\K_{\e e}})^D} =
\lim_{f_{\e}\to 1} \left(\frac{f_{\e}(\k+f_e)}{\k+f_{\e}f_e}\right)^D=1
\label{limNe}
\eeq

To start with, let us consider the simplest example, i.e. $\beta=0$,
which means $\EP(a^\dagger)=1$ in (\ref{ansatz}) and define the
candidate for the tachyon wavefunction. We will denote
$\hat\varphi_{e}(1,\t,p)$ by $\hat\varphi_{e}(\t,p)$ or simply by
$\hat\varphi_{e}$. We find that (\ref{genstar1}) takes the following
form
\beq
_1\langle \hat\Xi|_2\langle \hat \varphi_{e} |V_3\rangle=\lim_{\e\to 1}
{}_1\langle \hat\Xi_\e|_2\langle \hat \varphi_{e} |V_3\rangle=
\exp{\left[-\t\,a^\dagger p - \frac12G_1\,p^2\right]}
|\hat \Xi_{e}\rangle \, {\rm e}^{ipx}
\eeq
where $\t$ is a solution to
\beq
\t= \v_0-\v_+ +(V^{12},V^{21})\, \hat\K_{1e}^{-1}\,\hat\CS_{1e}\,\left(
\matrix{\v_+-\v_-\cr \v_--\v_0}\right) + (V^{12},V^{21})\, \hat\K_{1e}^{-1}\,
C\,\left(\matrix{0\cr \t}\right)\label{te1}
\eeq
and
\be
&&G_1 = 2V_{00} + (\v_+-\v_-,\v_--\v_0) \,\hat\K_{1e}^{-1} \,
\hat\CS _{1e} \left( \matrix{\v_+-\v_-\cr \v_--\v_0\cr}\right)\0\\
&&~~~~~~~~+2 (\v_+-\v_-,\v_--\v_0)\,\hat\K_{1e}^{-1}\, C
\left(\matrix{0\cr \t}\right)
+(0,\t) C\,\V  \,\hat\K_{1e}^{-1}\,C
\left(\matrix{0\cr \t}\right)\label{Ge1}
\ee
where $\hat\K_{1e}$ and $\hat\CS_{1e}$ equal $\hat\K_{\e e}$ and
$\hat\CS_{\e e}$ when $\e=1$, respectively.

If we repeat the same derivation for the other star product, we find
\beq
_1\langle \hat \varphi_{e}  |_2\langle  \hat\Xi |V_3\rangle=\lim_{\e\to 1}
{}_1\langle \hat\varphi_e|_2\langle \hat \Xi_\e|V_3\rangle=
\exp{\left[-\t\,a^\dagger p - \frac12 G_2\, p^2\right]}
|\hat \Xi_{e}\rangle \, {\rm e}^{ipx}
\eeq
where, this time, $\t$ is a solution to
\be
&&\t= \v_0-\v_- -(V^{12},V^{21})\,\hat \K_{e1}^{-1}\,\hat\CS_{e1}\,\left(
\matrix{\v_0-\v_+\cr \v_+-\v_-}\right) + (V^{12},V^{21})\, \hat\K_{e1}^{-1}\,
C\,\left(\matrix{\t\cr 0}\right)\label{te2}
\ee
and
\be
&& G_2 = 2V_{00} + (\v_0-\v_+,\v_+-\v_-) \,\hat\K_{e1}^{-1}\,\hat\CS_{e1}\,
\left(\matrix{\v_0-\v_+\cr \v_+-\v_-\cr}\right)\0\\
&&~~~~~~~~-2 (\v_0-\v_+,\v_+-\v_-)\,\hat\K_{e1}^{-1}\,
C \left(\matrix{\t\cr 0}\right)
+(\t,0) C\,\V  \,\hat\K_{e1}^{-1}\,C\left(\matrix{\t\cr 0}\right)\label{Ge2}
\ee
where $\hat\K_{e1}$ and $ \hat\CS_{e1}$ equal $\hat\K_{e\e }$ and
$\hat\CS_{e\e}$ when $\e=1$, respectively.

The two couples of expressions (\ref{te1},\ref{te2}) and
(\ref{Ge1},\ref{Ge2}) are formally different. Of course they must give
rise to the same result. If we require twist invariance
for $\t$, i.e. $C\t=\t$, it is easy to see that the two couples of
equations collapse to a single one. However, for reasons that will become
clear later on, we will not require twist invariance for $\t$
(see section 5.3 for more comments on this point). This is why we
wrote the two couples of equations explicitly. In general, therefore,
$\t=\t_++\t_-$. Hermiticity of the string field requires that
$C\t=\t^*$, i.e. $\t_+^*=\t_+$ and $\t^*_-=-\t_-$.

We remark now that, if the above equations have a nontrivial solution for
$\t$ and
\beq
e^{- \frac12 G p^2}=\frac 12,\label{12cond}
\eeq
where $G=G_1=G_2$, then $|\hat\varphi_e\rangle$ is a solution to the LEOM
(\ref{EOMlm}).

We also notice, for future use, that for a state of the general form
(\ref{ansatz}) to satisfy the LEOM, the equation for $\t$ and $G$ remain
the same.
The presence of a polynomial $\EP(a^\dagger)$ does not affect
the exponents, but only implies new conditions for the parameters
in $\EP(a^\dagger)$ (see below).

\newsection{Solution for $\t$ and $G$ }

In this section we study the solutions to eqs.(\ref{te1},\ref{te2})
and evaluate $G$.
Since, due to the structure of these equations, {\it a priori} one
cannot exclude the possibility of a singularity in $1-\e$, we
insert $\e$ at the right places and take the limit $\e\to 1$ on the
solution.
\subsection{The solutions for $\t$}
Let us see first
the relation between these two equations. We write $\t=\t_+ +\t_-$,
where $C\t_\pm = \pm \t_\pm$ and apply $C$ to (\ref{te1}). Keeping track of the
$\e$ dependence, we obtain
\beq
\t_+-\t_-= \v_0-\v_- +(X_+,X_-)\, \hat\K_{\e e}^{-1}\,\hat\T_{\e e}\,\left(
\matrix{\v_--\v_+\cr \v_+-\v_0}\right) + (X_+,X_-)\, \hat\K_{\e e}^{-1}\,
\left(\matrix{0\cr \t_+-\t_-}\right)\label{te1'}
\eeq
Doing the same with (\ref{te2}) we get
\beq
\t_+-\t_-= \v_0-\v_+ -(X_+,X_- )\, \hat\K_{e\e}^{-1}\,\hat\T_{e\e}\,\left(
\matrix{\v_0-\v_-\cr \v_--\v_+}\right) + (  X_+,X_- )\, \hat\K_{e\e}^{-1}\,
\left(\matrix{\t_+-\t_-\cr 0\cr}\right)\label{te2'}
\eeq
Next we introduce the operator $\sigma C$, where
$\sigma =\left(\matrix{0&1\cr 1&0\cr}\right)$. We have
\be
(\sigma C)^2=1,\quad\quad (\sigma C)\,\hat \K_{e\e}^{-1}\,(\sigma C)=
 \hat \K_{\e e}^{-1},
\quad\quad (\sigma C)\,\hat\T_{e\e}\,(\sigma C)= \hat\T_{\e e}
\ee
Therefore, by suitably inserting $(\sigma C)^2$ in (\ref{te2'}),
applying the above
transformations and applying $C$ to the resulting equation we find
\beq
\t_++\t_-= \v_0-\v_- +(X_+,X_-)\, \hat \K_{\e e}^{-1}\,\hat\T_{\e e}\,\left(
\matrix{\v_--\v_+\cr \v_+-\v_0}\right) + (X_+,X_-)\, \hat \K_{\e e}^{-1}\,
\left(\matrix{0\cr \t_++\t_-}\right)\label{te2''}
\eeq
Taking the sum and the difference of (\ref{te1'}) and (\ref{te2''})
we find separate equations for $\t_+$ and $\t_-$:
\be
&&\t_+= \v_0-\v_- +(X_+,X_-)\, \hat \K_{\e e}^{-1}\,\hat\T_{\e e}\,\left(
\matrix{\v_--\v_+\cr \v_+-\v_0}\right) + (X_+,X_-)\, \hat \K_{\e e}^{-1}\,
\left(\matrix{0\cr \t_+}\right)\label{t+}\\
&&\t_-=   (X_+,X_-)\, \hat \K_{\e e}^{-1}\,
\left(\matrix{0\cr \t_-}\right)\label{t-}
\ee

Now we have to solve these two equations. The rather lenghty calculations
are left for Appendix B. From the results therein one can see that,
for $\e=1$ and setting $\t_+=\t_0+\t_\a$, the first equation reduces to
\be
&&\t_0= 3\frac{T^2-T+1}{T+1}\v_0
\label{t0}\\
&&\left[1-\frac 1{\k+f_e}(|\xi\rangle+|C\xi\rangle)
\langle\xi|\frac{f_e+T}{1-T^2}\right] |\t_\a\rangle = 0
\label{redT+}
\ee
where $\t_0$ is the result obtained in \cite{HKw} (multiplied by $\sqrt{2}$).
It is easy to see that (\ref{redT+}) has the general solution
\beq
\t_\a = \a\, \langle\xi|\frac 1{T+1}|\t_0\rangle\,
(1+C)\xi  \label{genericsol}
\eeq
for any number $\a$. The factor $\langle\xi|\frac 1{T+1}|\t_0\rangle$
has been introduced for later convenience.

As for eq.(\ref{t-}) for $\e=1$ it has a nontrivial solution
\beq
\t_-=\b (1-C)\xi \label{solt-}
\eeq
with arbitrary $\b$.
This solution turns out to have an important role (see below).
In conclusion we can say that at $\e=1$ the solution for $\t$ can be written as
\beq
\t=\t_0 + \a\, \langle\xi|\frac 1{T+1}|\t_0\rangle\,
(1+C)\xi+\beta (1-C)\xi \label{solt}
\eeq
for arbitrary constants $\a$ and $\b$.

\subsection{Calculation of $G$.}

Once again, in order to compute $G$, we reintroduce the deformation
parameter $\e$ as in the previous section (see Appendix B).
We rewrite eqs.(\ref{Ge1},\ref{Ge2}) as follows
\be
&&G_1 = 2V_{00} + (\v_+-\v_-,\v_--\v_0) \,\hat\K_{\e e}^{-1}\,
\hat\T_{\e e}\, \left(
\matrix{\v_--\v_+\cr \v_+-\v_0\cr}\right)\label{G1}\\
&&~~~~~~~~+2 (\v_+-\v_-,\v_--\v_0)\,\hat\K_{\e e}^{-1}\,
\left(\matrix{0\cr \t_+-\t_-}\right)
+(0,\t_++\t_-) \,\M  \,\hat\K_{\e e}^{-1}\,
\left(\matrix{0\cr \t_+-\t_-}\right)\0
\ee
and
\be
&& G_2 = 2V_{00} + (\v_0-\v_+,\v_+-\v_-) \,\hat\K_{e\e}^{-1}\,\hat\T_{e\e}\,
\left(\matrix{\v_0-\v_-\cr \v_--\v_+\cr}\right)\label{G2}\\
&&~~~~~~~~-2 (\v_0-\v_+,\v_+-\v_-)\,\hat\K_{e\e}^{-1}\,
\left(\matrix{\t_+-\t_-\cr 0}\right)
+(\t_++\t_-,0) \,\M  \,\hat\K_{e\e}^{-1}\,
\left(\matrix{\t_+-\t_-\cr 0}\right)\0
\ee
Using (\ref{solt}) we obtain
\beq\label{G1calc}
G_1 = G_0 - 2\,(f_\e-1) \frac{\k+f_e}{\k+f_\e f_e} \left[ \a
(1-\k\a) \left(\langle\t_0|\frac 1{1+T}|\xi\rangle\right)^2
+ \b \left( \k\b + \langle\t_0|\frac 1{1+T}|\xi\rangle \right)
\right]
\eeq
and
\beq\label{G2calc}
G_2 = G_0 - 2\,(f_\e-1) \frac{\k+f_e}{\k+f_\e f_e} \left[ \a
(1-\k\a) \left(\langle\t_0|\frac 1{1+T}|\xi\rangle\right)^2
+ \b \left( \k\b - \langle\t_0|\frac 1{1+T}|\xi\rangle \right)
\right]
\eeq
Therefore, for $\e=1$ we obtain $G_1=G_2=G_0$.
Naive manipulations of the relevant formulas lead to the result $G_0=0$.
However $G_0$ contains two divergent terms, which need to be regularized.
As shown by Hata et al. \cite{HKw, HM1, HM2}, using level truncation one
obtains\footnote{Our
definitions for $\t$ and $G$ differ from those in \cite{HKw} by
factors of $\sqrt2$ and 2, respectively, see Appendix A.} $G_0= 2 \ln \,2$.

\newsection{The tachyon and vector excitations}

After a long preparation we are now ready to start the analysis of the
fluctuations around the dressed sliver.

\subsection{The tachyon excitation}

From the results of the previous section it follows that string fields of the
form
\beq \label{etachyon}
|\hat\varphi_e(\t,p)\rangle\,=\,\N_e \, \exp{\Big(-\sum_{n\geq 1}
t_n a_n^{\mu\dagger}\hat p_\mu \Big)}
|\hat \Xi_e\rangle\,  {\rm e}^{ipx}
\eeq
with $\t$ as in (\ref{solt}), satisfy the LEOM when the momentum fulfills
the mass-shell condition $m^2=-p^2=-1$. This solution depends on three
arbitrary parameters $e$, $\a$ and $\b$. Eventually we shall see that
in fact we have to set $e=1$. As we will see, the other two parameters
never enter the evaluation of physical quantities. There is one
more question.
We expect the tachyon to be represented by a twist-even state,
and we already noticed that (\ref{etachyon}) does not have definite
twist parity. We will see at the end of this section how to settle
this problem.

\subsection{The vector excitation}

Fluctuations other than the tachyon can be obtained by considering
nontrivial polynomials in eq.(\ref{ansatz}). The polynomial will consist of
sum of monomials of the type
\beq
d^{\mu_1\ldots\mu_p} \langle\zeta_1 a_{\mu_1}^\dagger\rangle\,\ldots\,
\langle\zeta_p
a_{\mu_p}^\dagger\rangle  \label{monom}
\eeq
where $\langle\zeta_i a^{\mu_i\dagger}\rangle =
\sum_{n>0}\zeta_{in} a_n^{\mu_i\dagger}$.
As it turns out the $\e$--dependence is trivial as far as higher fluctuations are
concerned, therefore we drop it throughout.

Let us find the level one state, corresponding to the
massless vector. We start with the following ansatz for the
matter part
\beq
|\hat\varphi_{e,v}(d^\mu,\t,p)\rangle =
\N_v\N_e d^\mu \langle(1-C) \zeta a_\mu^\dagger\rangle\,
e^{-\sum_{n\geq 1} t_n a_n^{\mu\dagger}\hat p_\mu}
|\hat \Xi_e \rangle{\rm e}^{ipx}=
\N_v d^\mu \langle(1-C) \zeta a_\mu^\dagger\rangle
|\hat \varphi_e(\t,p)\rangle
 \label{vectoransatz}
\eeq
with $\rho_2\zeta=\zeta$ and $\rho_1\zeta=0$.

Using the results of Appendices D and E we obtain
\be
&&|\hat\varphi_{e,v} \rangle \,* \,|\hat \Xi\rangle +
|\hat \Xi\rangle \,*\,|\hat\varphi_{e,v} \rangle
= e^{-\frac12 G\,p^2} \left[
d^\mu \langle(1-C) \zeta a_\mu^\dagger\rangle \,+ \right.
\0 \\
&& ~~~~ \left. + \,
\frac{1}{\k+f_e} \langle\xi|\frac{f_e+T}{1-T^2}|\zeta\rangle
d^\mu \langle(1-C) \xi a_\mu^\dagger\rangle
+ 2\b\, (p\cdot d) \, \langle\xi|\frac{\k-T}{1-T^2}|\zeta\rangle
\right] \N_v \, |\hat \varphi_e(\t,p)\rangle
\label{vectorsol}
\ee
From this result we see that in order to satisfy the
LEOM we have to assume that $p^2=0$ and to impose the transversality
condition
\beq
 p \cdot d =0 \label{transvers}
\eeq Therefore we recover the massless vector state with the
correct transversality condition. This result is independent of
the value of $e$. In order to satisfy the LEOM we also have to
impose \beq \bra{\xi} \frac
{f_e+T}{1-T^2}\ket{\zeta}=0\label{condvector} \eeq This is to be
understood as a condition on the vector $\zeta$ and as such it is
easy to comply with it. For reasons that will become clear later,
eventually we will set $e=1$. In this case (\ref{condvector})
becomes simply \be \bra{\xi} \frac 1{1-T}\ket{\zeta}=0\0 \ee which
is the condition of orthogonality to the extra eigenvector(s) of
the dressed sliver (\ref{extra}). To conclude we remark that
dressing is essential in order to obtain the transversality
condition.

\subsection{Twist parity}
Let us see how to implement the requirement that the solutions to the LEOM have
definite twist parity.

The twist operator $\Omega$ acts on Fock-space mode operators as
\begin{equation}
\Omega (a,b,c)_n \Omega = (-1)^n (a,b,c)_n = (Ca,Cb,Cc)_n
\end{equation}
and satisfies $\Omega|p\rangle=|p\rangle$. Acting on the state
(\ref{etachyon}) one obtains
\begin{equation}\label{Comega}
\Omega \, |\hat\varphi_e(\t,p)\rangle \,=\,
|\hat\varphi_e(C\t,p)\rangle
\end{equation}
It follows that the states we have constructed (tachyon and vector)
in general are not eigenstates of $\Omega$. This is in contrast to the
expectation that the tachyon should be twist-even, the vector twist-odd, etc.
Now, from the properties of the VSFT action it follows that if some
string field $|\Psi\rangle$ is a solution to the LEOM, then
$\Omega|\Psi\rangle$ is also a solution. Using this we can define
twist parity eigenstates in the following way:
\begin{itemize}
\item \textbf{Tachyon} \ ($\Omega=1$) :
\begin{equation} \label{tachim}
|\hat\varphi_t(\t,p)\rangle \,=\, \frac 12\,
(1+\Omega) \, |\hat\varphi_e(\t,p)\rangle
\end{equation}
\item \textbf{Vector} \ ($\Omega=-1$) :
\begin{equation} \label{vcond}
|\hat\varphi_v(d^\mu,\t,p)\rangle \,=\,  \frac 12\,
 (1-\Omega)|\hat\varphi_{e,v}(d^\mu,\t,p)\rangle
\end{equation}
where $|\hat\varphi_{e,v}(d^\mu,\t,p)\rangle$ is given in
(\ref{vectoransatz}).
\end{itemize}
This naturally generalizes to higher states.

At the end of section 3 we noted that the string field
$\ket{\hat\varphi_e(\t,p)}$ will satisfy the reality condition if
$C\t=\t^*$, which when applied to (\ref{solt}) implies that $\a$
is real and $\b$ is pure imaginary. Consequently, the same applies
to parity eigenstates (\ref{tachim}) and (\ref{vcond}). But, now
it is easy to see that there is another possibility:
(\ref{tachim}) satisfies the reality condition also when $\b$ is
real, and the same is true for (\ref{vcond}) when multiplied by
$i$. In short, we take $\a$ to be real while $\b$ can be either
real or pure imaginary.\\
Having seen how to implement the correct twist parity on the
states, it should be mentioned that, in computing physical
observables like masses and amplitudes, one could  use the twist
asymmetric states (\ref{Comega}) since the twist violating part
(which is controlled by the parameter $\beta$) never enters in
such observables, see also section 10.

\newsection{Probing the $k\sim 0$ region}

Level truncation is a natural regularization in the SFT context.
It permits many numerical computations, but it is very unwieldy
if one wants to derive analytical results, the lack of analytical
control being related to the impossibility of using the analytical
machinery of the continuous basis. This is true in particular for
the region around $k=0$, i.e. the string midpoint region,
which turns out to be crucial for higher level excitations.
In this section we therefore introduce an analytic surrogate
of level truncation, at least as far as the $k\sim 0$ region is concerned.
It consists of a regulator which mimics the level truncation
by regulating the singularities arising when the $k\sim0$ region is
probed but has the good feature of being
defined on the continuous basis (hence permitting analytical control).

To this end the crucial issue is the eigenvalues distribution at
$k\sim 0$. As proved in \cite{RSZ5} this distribution is
divergent, but can be regularized in large--$L$ level truncation
\beq \rho(k)=\frac{\ln L}{2\pi} + \rho_{fin}(k)\0 \eeq the
quantity $\rho_{fin}(k)$ is responsible for finite contributions
which are relevant for large $k$, see \cite{belov}, but it will
play no role in the sequel. The eigenvectors of the $k$--basis
have infinite norm due to the continuous orthonormality condition
\beq \bra k k'\rangle=\delta(k-k') \eeq Large--$L$ level
regularization suggests that their norm is given by\footnote{Again
finite corrections are neglected, as they are not important for
our purposes.} \beq \bra k k\rangle=\delta(0)=\frac{\ln L}{2\pi}
\eeq Consider now the following half (right) string vector in the
$k$--basis \beq\label{epsi} \ket\eta =\frac 1\eta
\int_{\frac\eta2}^{\frac{3\eta}{2}} dk \ket k, \quad\quad \eta>0
\eeq The norm of this vector is easily computed to be
\beq\label{normepsi} \bra\eta\eta\rangle=\frac1\eta \eeq From this
we define a twist--even and a twist--odd vector as follows
\be\label{epsipm}
\ket{\eta_+} &=&\frac1{\sqrt{2}} \left( \ket\eta+C\ket\eta\right)\0\\
\ket{\eta_-} &=&\frac1{\sqrt{2}} \left( \ket\eta-C\ket\eta\right)
\ee
Their norm is given by
\beq\label{normepsipm}
\bra{\eta_-}\eta_-\rangle=\bra{\eta_+}\eta_+\rangle=\frac{1}{\eta}
\eeq
These two vectors are the basis of our regularization.
In the limit $\eta\to 0^+$ they collapse to the midpoint $k=0$, and keeping
track of the powers of $\eta$ will allow us to give an unambiguous meaning
to the objects we are interested in.

Our first aim is to show that this procedure is inspired by and very close to
the level truncation. To this end let us expand these two vectors in the
oscillator basis $\ket n$. Using
\be
\bra n k\rangle=\sqrt{\frac{n k}{2\sinh\frac{\pi k}{2}}}\oint\frac{dz}{2\pi i}
 \frac{1}{z^{n+1}}\frac{1}{k} \left(1-\exp(-k\tan^{-1}z)\right)\0
\ee
a term by term integration yields
\be\label{nepsi}
\bra n\eta_-\rangle &=& \sqrt{\frac{{2}}{{\pi}}}\left(1,0,-\frac{1}{\sqrt{3}},0,
\frac{1}{\sqrt{5}},0,...\right)+O(\eta^2)\0\\
\bra n\eta_+\rangle &=& -\frac{\eta}{\sqrt{2\pi}}\left(0,\sqrt{2},0,-\frac43,0,
\frac{23}{15}\sqrt{\frac23},0,...\right)+O(\eta^2)
\ee
The first vector is therefore the usual $\ket{k=0}$ twist--odd vector, while
every component of the second vanishes in the limit $\eta\to 0$.
The latter is ($-\eta\sqrt{2/\pi}$) times
the $K^2=0$ twist--even vector Rastelli, Sen and Zwiebach found in
\cite{RSZ5}, that is
\be\label{0-+}
&& |0_-\rangle=\lim_{\eta\to 0^+}\ket{\eta_-} = \sqrt{\frac 2\pi} \,
\ket{v_{RSZ}^-}\label{v+}\\
&& |0_+\rangle=\lim_{\eta\to 0^+}\ket{\eta_+} =
-\eta\sqrt{\frac 2\pi} \, \ket{v_{RSZ}^+}\label{v-}
\ee
It is important to note that although the twist--even vector
$\ket{0_+}$ is vanishing, due to
(\ref{normepsipm}), it has the same infinite norm as $\ket{0_-}$. Like
all the vectors which form the continuous basis, this
vector does not belong either to the Fock space, but, unlike
all other $\ket k$'s,  it has vanishing overlap with all oscillators
\be\label{nezero+}
\bra n 0_+\rangle=\lim_{\eta \to 0}\bra n \eta_+\rangle=0
\ee
Nevertheless,
as we will see in the sequel, it is crucial for the consistency
of the $*$--algebra and, moreover,
for accommodating the complete open string D--brane spectrum in the VSFT
approach.

At this stage it should be clear that the $\eta$ parameter plays the role
of an
effective large L truncation of the continuous basis, and that
$\ket{\eta_-}$
represents the eigenvector relative to the smallest eigenvalue of $T$ at level
$L(\eta)$, which is always twist--odd.
From \cite{HM2} we expect the first eigenvector to be located at
$k=\frac{\pi}{\log L}$. This suggests that one should make the identification
\beq\label{epsiL}
\eta=\frac{\pi}{\log L}
\eeq
We can verify this assertion by checking that
\be
\bra{0_-} 0_-\rangle=\bra{0_+} 0_+\rangle\0
\ee
Using (\ref{0-+}), this gives
\be
\eta \,=\, \sqrt{\frac{\bra{v_{RSZ}^-} v_{RSZ}^-\rangle}{\bra{v_{RSZ}^+}
 v_{RSZ}^+\rangle}}\label{etaexp}
\ee
Computing the difference between the RHS of (\ref{epsiL})
and the RHS of (\ref{etaexp}) in level truncation we find that  it
becomes smaller and smaller as $L\to \infty$.
For example at
$L=1000$ we have $\frac{\pi}{\log L}\sim 0.45479$ (not very near 0!) and
such a difference is $-0.03082$, while at $L=10000$ we have
$0.34109$ and $-0.01040$, respectively, which is a $3\%$ agreement.
Proceeding further with the level it is easy to verify that the
agreement improves\footnote{This simple example should warn the reader on how level
truncation is slow in probing the midpoint $k=0$.}.

We have therefore succeeded in relating our regularization parameter
$\eta$ to the cutoff $L$. With some abuse of language we will call the
previous empirical set of rules $\eta$--regularization. Now we are
going to show that some ambiguities that used to plague the string
midpoint analysis, within this regularization scheme are naturally
resolved. We are interested, in particular, in the action of the half
string projectors $\rho_{1,2}$ on the midpoint modes $\ket{0_\pm}$. By
using the $\eta$--regularization (\ref{epsipm}) we simply get
\be\label{rho0pm}
&&\rho_1\ket{0_\pm}=\frac12\ket{0_\pm}+\frac12\ket{0_\mp}\0\\
&&\rho_2\ket{0_\pm}=\frac12\ket{0_\pm}-\frac12\ket{0_\mp}\0\\
&&(\rho_1-\rho_2)\ket{0_\pm}=\ket{0_\mp}
\ee
If we contract this result with any Fock space vector $\bra n$, we recover
the result of \cite{HK} that the $\rho$ projectors have
$\frac12$ eigenvalue
at $k=0$. The latter assertion is however, by itself, not free from
ambiguities and/or associativity inconsistencies if we do not
want to give up the properties (\ref{proj12}). For example, a naive
manipulation leads to
\beq
0=(\rho_1\rho_2)\ket{0_-}\neq\rho_1(\rho_2\ket{0_-})=\frac{1}{4}\ket{0_-}
\eeq
On the contrary, with our regularization it is very easy to check that
\beq
0=(\rho_1\rho_2)\ket{0_\pm}=\rho_1(\rho_2\ket{0_\pm})=0
\eeq
which is definitely non--ambiguous.
Other remarkable inconsistencies which would
arise using the same kind of naive manipulations would be
\be
\frac{1}{2}\ket{0_-}=(\rho_{1,2}\rho_{1,2})\ket{0_-}\neq\rho_{1,2}
(\rho_{1,2}\ket{0_-})=\frac14\ket{0_-}\0\\
\ket{0_-}=(\rho_1-\rho_2)^2\ket{0_-}\neq(\rho_1-\rho_2)\left((\rho_1-\rho_2)
\ket{0_-}\right)=0
\ee
It is easy to check that, with our regularization, this anomaly
disappears and all the properties (\ref{proj12}) are preserved even at $k=0$.
The crucial move was to introduce
an extra twist--even midpoint vector which vanishes in the Fock space, but has
nevertheless infinite norm. We will see in the sequel how this vanishing vector
is important for the construction of open string states on the dressed sliver.
For the time being we only point out that
the vector $\ket{0_+}$ cannot create string excitations when contracted with
oscillators since, see (\ref{nezero+}),
\beq\label{vanishstate}
\bra{0_+}a^{\dagger}\rangle |state \rangle=
\lim_{\eta \to 0}\sum_n a_n^\dagger\, \bra n \eta_+\rangle|state\rangle=0
\eeq
vanishes. However we can excite Fock space states
if, in  $\eta$--regularization, we consider the vector
\beq\label{O+fin}
\lim_{\eta\to 0^+}\frac1\eta\ket{\eta_+}\sim\ket{v_{RSZ}^+}
\eeq
From (\ref{nepsi}) it is clear  that this vector has finite overlap with any
Fock space vector. We will see that this vector plays a fundamental role
in the construction of cohomologically non--trivial open string states.
The vector $\ket{0_+}$ can also contribute to  matrix elements involving
vectors that are finite at the midpoint (hence out of the Fock space)
like the ``bare tachyon'' $\bra{t_0}$. For example the following relations hold
in $\eta$--regularization
\be\label{rel0+}
\bra{t_0}0_+\rangle &=& \sqrt{2} \, t_0(0) + O(\eta)\\
\bra{t_0}\frac1{1+T}\ket{0_+} &=& \ln3 \, \frac{2\sqrt{2}}{\pi} \,
t_0(0) \frac1\eta + O(1)
\ee

In the sequel we will see that, using $\eta$--regularization, all the
divergent brackets that appear in computing solutions to the LEOM can be
explicitly evaluated in terms of some (regularization dependent) function
of $\eta$. We will comment a posteriori on the regularization independence
of our final and physical results.

\newsection{Higher level solutions to LEOM}

In the canonical quantization of string theory the tower of massive states
is constructed by applying monomials of creation operators on the Fock vacuum.
In order for the state to have a definite mass one selects all the monomials
of the same level and takes a linear combination thereof, with tensorial
coefficients which are generically referred to as polarizations. The latter are
not completely free, but must satisfy some constraints, the Virasoro constraints.
The construction of analogous states in VSFT proceeds differently. Although
we will keep talking about level $n$ solutions in order to relate our results
with the familiar ones, the
level is not the right issue here, because in VSFT we do not have any explicit
realization of the $L_0$ Virasoro generator. The most general level
$n$ state we
will consider will take the form
\beq
\ket{\hat \varphi(\theta,n,\t,p)}\equiv\ket{\hat \varphi(\theta_1,...,
\theta_n,\t,p)} =
\sum_{i=1}^n
\theta^{\mu_1\ldots\mu_i}_i \langle a_{\mu_1}^\dagger \zeta^{(i)}_1\rangle
\ldots \langle a_{\mu_i}^\dagger \zeta^{(i)}_i\rangle \ket{\hat\varphi(\t,p)}
\label{levelnst}
\eeq
in analogy with the canonical quantization construction, but without imposing
any level restriction.
As we shall see below, the request that the state (\ref{levelnst}) satisfy the LEOM
will be sufficient to select a definite mass and impose the appropriate
Virasoro constraints on the polarizations $\theta_i$.

\subsection{Level 2}
The level 2 state in canonical quantization is
\beq
(h_\mu \a_2^{\mu\dagger}+ \lambda_{\mu\nu} \a_1^{\mu\dagger}
\a_1^{\nu\dagger})\ket{0}e^{ipx}\;.
\label{level2}
\eeq
The Virasoro constraints require that $p^2=-1$ and
\beq
2\sqrt2 \, h^\mu p_\mu + \lambda_\mu {}^{\mu} =0 \,, \quad\quad
h_\mu + \sqrt2 \, \lambda_{\mu\nu} p^\nu =0\label{Vir2}
\eeq
In view of the forthcoming VSFT construction it is important to notice that
there is a certain arbitrariness in these formulas. One can rewrite them for instance
as follows
\beq
2\sqrt2 \, g^\mu p_\mu + a \theta_\mu {}^{\mu} =0 \,, \quad\quad
b g_\mu + \sqrt2 \, \theta_{\mu\nu} p^\nu =0
\label{Vir2'}
\eeq
with $a$ and $b$ arbitrary (non--vanishing) constants, and $h,\lambda$ related to
$g,\theta$ as follows
\be
h_\mu =A g_\mu + B(p\cdot g) p_\mu,\quad\quad
\lambda_{\mu\nu}= C \theta_{\mu\nu} + D(p_\mu p^\rho \theta_{\rho\nu}
+ p_\nu p^\rho \theta_{\mu\rho})\label{redef2}
\ee
Using the mass--shell condition it is easy to show that this simply requires
\be
A =  \frac{b}{2} \, \frac{3ab+2}{ab-1}\,D \,, \quad\quad
B = b\,D\,, \quad\quad C = \frac{5}{2}\,\frac{ab}{ab-1}\,D \0
\ee

According to the level $n$ ansatz (\ref{levelnst}) the candidate to represent
a level 2 state is
\beq
\ket{\hat \varphi(\theta,2,\t,p)}\equiv\ket{\hat \varphi(\theta_1,
\theta_2,\t,p)} =\theta^{\mu_1}_1 \langle a_{\mu_1}^\dagger\zeta^{(1)}_1\rangle
  \ket{\hat\varphi(\t,p)} +
\theta^{\mu_1\mu_2}_2 \langle a_{\mu_1}^\dagger \zeta^{(2)}_1\rangle
\langle a_{\mu_2}^\dagger \zeta^{(2)}_2\rangle \ket{\hat\varphi(\t,p)}
\label{level2st'}
\eeq
This ansatz has to be made more precise by specifying the vectors $\ket{\zeta_j^{(i)}}$.
For generic vectors we do not get any on--shell open string state. In fact,
on the basis of our attempts, it seems that
only if the vectors $\ket{\zeta_j^{(i)}}$ probe the string midpoint will
(\ref{level2st'}) be a cohomologically non--trivial
solution to the LEOM. Therefore we make the choice
$\ket{\zeta_j^{(i)}}\sim \ket{0_\pm}$; the latter states were introduced in
the previous section and were designed to resolve the singularity
at $k=0$. But we must be more precise:  the factors in front of
$\lim_{\eta\to 0^+}\ket{\eta_\pm}$ play also a fundamental role and we must specify
them. In summary, our ansatz will be
 \beq
\ket{\hat \varphi(g,\theta,\t,p)} =g^{\mu} \langle a_{\mu}^\dagger\ket{s_+}
  \ket{\hat\varphi(\t,p)} +
\theta^{\mu\nu} \langle a_{\mu}^\dagger \ket{\zeta_-}
\langle a_{\nu}^\dagger \ket{ \zeta_-} \ket{\hat\varphi(\t,p)}
\label{level2st}
\eeq
where $\ket{s_+}= \lim_{\eta\to 0^+} \ket{\eta_+}s(\eta)$, $\ket{ \zeta_-} =
\lim_{\eta\to 0^+}  \ket{\eta_-}\zeta(\eta)$,
and, near $\eta=0$,
\beq
s(\eta) = \frac{s_{-1}}{\eta} +s_0+s_1\eta+\ldots,\quad\quad
\zeta(\eta) = \zeta_0+ \zeta_1\eta+ \zeta_2\eta^2+\ldots\label{expan}
\eeq
As a consequence we have (see (\ref{v+},\ref{v-}))
\be
&&\langle a_{\mu}^\dagger\ket{s_+} = - \sqrt{\frac 2{\pi}}
\langle a_{\mu}^\dagger\ket{v^+_{RSZ}}\,(s_{-1} + s_0\eta
+ s_1\eta^2+\ldots) \label{as+}\\
&&\langle a_{\mu}^\dagger \ket{\zeta_-} = \sqrt{\frac 2{\pi}}
\langle a_{\mu}^\dagger\ket{v^-_{RSZ}}
\, (\zeta_0+ \zeta_1\eta+ \zeta_2\eta^2+\ldots)\label{azeta-}
\ee
These are well--defined expressions and
it would seem that the terms proportional to $\eta,\eta^2$ play no role in the limit
$\eta\to 0$. However this is not the case because the star product with the dressed
sliver will take them back into the game. Only terms of order $\eta^3$ and
higher will not play any role and can be disregarded.

It is time to pass to the explicit calculation of the LEOM. We have to find the
conditions under which
\be
 \ket{\hat \varphi(g,\theta,\t,p)} * \ket{\hat\Xi} + \ket{\hat\Xi}*
\ket{\hat \varphi(g,\theta,\t,p)} = \ket{\hat \varphi(g,\theta,\t,p)} \label{leom2}
\ee
The star products in (\ref{leom2}) yield
cumbersome formulas. In order not to clog our exposition with them we defer a
full treatment to Appendix F, and use a technical simplification:  we
assume that the function $\xi(k)$, which represents
the dressing vector $\xi$ in the $k$--basis and which is non-vanishing only for
negative $k$, is actually non--vanishing only for $k<k_0<0$ where $k_0$ is some
small but finite negative constant. The consequences of this simplification
will be commented upon in section 8. We can of course suppose that the regularization
parameter $2\eta<|k_0|$. As a consequence all the quantities appearing in
this computation which involve $\xi$ can be neglected. On the other hand
this restriction on the form of $\xi(k)$ does not imperil the properties
we have requested for $\xi$ in this and the previous paper \cite{BMP1}:
this point is further developed in section 8.
With this understanding we obtain
\be
&&\left( \theta^{\mu\nu} \langle a_{\mu}^\dagger \ket{\zeta_-}
\langle a_{\nu}^\dagger \ket{ \zeta_-} \ket{\hat\varphi(\t,p)}\right)*
\ket{\hat\Xi} + \ket{\hat\Xi}*\left(\theta^{\mu\nu}
\langle a_{\mu}^\dagger \ket{\zeta_-}
\langle a_{\nu}^\dagger \ket{ \zeta_-} \ket{\hat\varphi(\t,p)}\right)
\label{leom21}\\
&&= e^{-\frac 12 Gp^2}\left[\frac 12\theta^{\mu\nu} \,
\langle a_{\mu}^\dagger \ket{\zeta_-}\,
\langle a_{\nu}^\dagger \ket{ \zeta_-}+ 2\, \theta_\mu {}^\mu \,
\bra{\zeta_-} \frac T{1-T^2} \ket{\zeta_-}\right.\0\\
&&\left.\quad\quad\quad\quad\quad+\,2\, \theta^{\mu\nu}\,
\langle a_{\mu}^\dagger \ket{ \zeta_+} \,
p_\nu\, \EH_+ + 2  \theta^{\mu\nu} p_\mu p_\nu \EH_+^2 \right]
\ket{\hat\varphi(\t,p)} \0
\ee
where we have used $\ket{\zeta_+} = (\rho_1-\rho_2)\ket{\zeta_-}$, with
(\ref{vanishstate}), and
\be
&&\left(g^{\mu} \langle a_{\mu}^\dagger\ket{s_+}
\ket{\hat\varphi(\t,p)}\right)*\ket{\hat\Xi}+\ket{\hat\Xi}*
\left(g^{\mu} \langle a_{\mu}^\dagger\ket{s_+}
\ket{\hat\varphi(\t,p)}\right)
\label{leom22}\\
&&\quad\quad\quad=e^{-\frac 12 Gp^2}\left[g^{\mu}
\langle a_{\mu}^\dagger\ket{s_+} - p\cdot g \,\bra{t_0}
\frac 1{1+T} \ket {s_+}\right]\ket{\hat\varphi(\t,p)} \0
\ee
The quantity $\EH_+$ (see appendix F) is a complicated expression of
order $\eta^{-1}$,
$\,\bra{\zeta_-} \frac T{1-T^2}\ket{\zeta_-}$,
as well as $\,\bra{t_0}\frac 1{1+T} \ket {s_+}$, is of order $\eta^{-2}$,
while,
as we have already seen, $\langle a_{\mu}^\dagger \ket{ \zeta_+}$ is
of order $\eta$.

Now, from the first term in the RHS of eq.(\ref{leom21}) we see that the only
way to satisfy the LEOM is to set $e^{-\frac 12 Gp^2}=2$, i.e. $p^2=-1$, which
reproduces the
desired mass--shell condition. Next, in
(\ref{leom22}) we must split $g^{\mu} \langle a_{\mu}^\dagger\ket{s_+}$
(which is a finite term in $\eta$) in two halves. The first half
reconstructs the first term in the RHS of (\ref{level2st}), the second
half must annihilate the linear term in $a^\dagger$ in the RHS of
(\ref{leom21}): this is the only way this unwanted term can be
canceled. The latter operation on the other hand is only possible if
\be
g^\mu \sim \theta^{\mu\nu}p_\nu\label{vircon2}
\ee
Finally the remaining unwanted terms in the above equations
must cancel with one another order by order in $\eta$. Looking at the order
--2 in $\eta$, one easily realizes that the only way to implement such
cancelation is to require that
\be
\theta_\mu {}^\mu\sim \theta^{\mu\nu} p_\mu p_\nu \sim p\cdot g
\label{vircon1}
\ee
with nonvanishing proportionality constants.

Eqs.(\ref{vircon1},\ref{vircon2}) are not enough to conclude that the
level 2 Virasoro constraints (\ref{Vir2'}) are satisfied. However the
accurate analysis of Appendix F proves that this is the case. In
Appendix F it is also shown that the LEOM (\ref{leom2}) is exactly
satisfied together with the Virasoro constraints (\ref{Vir2'}),
provided some (not very restrictive) relations among the constants
$a,b,\zeta_0,\zeta_1,\zeta_2,s_{-1},s_0,s_1$ are satisfied. From the
analysis in Appendix F it is clear that the coefficients $a$ and $b$
are regularization dependent, but, in turn, $a$ and $b$ can be absorbed
via the redefinitions (\ref{redef2}).

\subsection{Level 3}

The level 3 state in canonical quantization is
\beq
(h_\mu \a_3^{\mu\dagger}+ \lambda_{\mu\nu} \a_2^{\mu\dagger}\a_1^{\nu\dagger}
+\chi_{\mu\nu\rho}\a_1^{\mu\dagger} \a_1^{\nu\dagger}  \a_1^{\rho\dagger}   )
\ket{0}e^{ipx},\label{level3}
\eeq
The Virasoro constraints require that $p^2=-2$ and
\be
3 h^\mu p_\mu + \,\sqrt{2}\, \lambda_\mu{}^\mu = 0\,, && \qquad
3 h_\mu + \sqrt2 \, \lambda_{\mu\nu} p^\nu = 0
\label{Vir3'a}\\
 \sqrt2 \,(2\, \lambda_{\nu\mu} p^\nu -\lambda_{\mu\nu} p^\nu)+
3 \chi_{\mu\nu}{}^{\nu} = 0 \,, && \qquad
\sqrt2 \, \lambda_{(\mu\nu)} + 3 \chi_{\mu\nu\rho} p^\rho = 0
\label{Vir3'b}
\ee
where $\lambda_{(\mu\nu)}$ is the symmetric part of $\lambda_{\mu\nu}$.
It can be seen that the first equation is a consequence of the other
three. It is however possible, as above, to redefine the polarizations
as shown in Appendix G. In terms of the new ones
$g_\mu,\omega_{\mu\nu},\theta_{\mu\nu\rho}$ the Virasoro constraints
become
\be
3\, x\, g^\mu p_\mu + \sqrt{2}\, \omega_\mu{}^\mu =0 \,, && \qquad
3\, g_\mu + \sqrt2\, y\,\omega_{\mu\nu} p^\nu = 0
\label{Vir3a}\\
 2\sqrt2 \,v\, \omega_{\nu\mu} p^\nu - \sqrt2 \,u\, \omega_{\mu\nu} p^\nu+
3\, \theta_{\mu\nu}{}^{\nu} = 0 \,, && \qquad
\sqrt2 \,\omega_{(\mu\nu)} + 3\, z\,\theta_{\mu\nu\rho} p^\rho = 0
\label{Vir3b}
\ee
It is now easy to verify that the first condition is a consequence
of the other three provided we set $x = \frac {z(2v-u)}{y}$. Therefore it
need not be verified separately. The remaining constants
$y,u,v,z$ are arbitrary non--vanishing ones. From the general form (\ref{levelnst}), we
select the following ansatz
\be
&&\ket{\hat \varphi(g,\omega,\theta,\t,p)}\label{level3st}\\
&&\quad\quad\quad =
\left(g^{\mu} \langle a_{\mu}^\dagger\ket{r_-}+
\omega ^{\mu\nu} \langle a_{\mu}^\dagger \ket{\zeta'_-}
\langle a_{\nu}^\dagger \ket{ \lambda_+} +
\theta^{\mu\nu\rho} \langle a_{\mu}^\dagger \ket{\zeta_-}
\langle a_{\nu}^\dagger \ket{\zeta_-} \langle a_{\rho}^\dagger
\ket{\zeta_-}\right)\ket{\hat\varphi(\t,p)}
\0
\ee
where $\ket{ r_-} =\lim_{\eta\to 0^+}  \ket{\eta_-}r(\eta)$,
and the same definition is understood for $\ket{ \zeta'_-}, \ket{ \zeta_-}$,
while $\ket{\lambda_+}= \lim_{\eta\to 0^+} \ket{\eta_+}\lambda(\eta)$.
Near $\eta=0$,
\beq
\lambda(\eta) = \frac{\l_{-1}}{\eta} +\l_0+\l_1\eta+\ldots,\quad\quad
\zeta(\eta) = \zeta_0+ \zeta_1\eta+ \zeta_2\eta^2+\ldots\label{expan'}
\eeq
$\zeta'(\eta)$ and $r(\eta)$ have an expansion similar to $\zeta(\eta)$.
Consequently, for the brackets inside (\ref{level3st}), expansions
similar to (\ref{as+}) and (\ref{azeta-}) hold.

The formulas involved in the evaluation of the linearized EOM are too large
to be written down here.
We can avoid such complications by introducing the simplifying assumption
of the previous subsection: we render the dressing vector contributions
evanescent in the limit $\eta\to 0$ so that we can simply avoid writing
them down. The resulting formulas are as follows:
\be
&&\left(\theta^{\mu\nu\rho} \langle a_{\mu}^\dagger \ket{\zeta_-}
\langle a_{\nu}^\dagger \ket{\zeta_-} \langle a_{\rho}^\dagger
\ket{\zeta_-}\ket{\hat\varphi(\t,p)}\right) *\ket{\hat\Xi}
+ \ket{\hat\Xi} * \left(\theta^{\mu\nu\rho} \langle a_{\mu}^\dagger
\ket{\zeta_-}
\langle a_{\nu}^\dagger \ket{\zeta_-} \langle a_{\rho}^\dagger
\ket{\zeta_-}\ket{\hat\varphi(\t,p)}\right) \0\\
&&= e^{-\frac 12 Gp^2}\left[ 3\, \theta_\mu {}^{\mu\rho} \,\bra{\zeta_-}
\frac T{1-T^2}\ket{\zeta_-}\langle a_\rho^\dagger\ket{\zeta_-}
+\theta^{\mu\nu\rho} \,\left(\frac 14\langle a_{\mu}^\dagger \ket{\zeta_-}\,
\langle a_{\nu}^\dagger \ket{ \zeta_-}
\langle a_{\rho}^\dagger \ket{ \zeta_-}\right.\right.\label{thetastar3}\\
&&\left. \left.\quad\quad\quad\quad+3\, \langle a_{\mu}^\dagger
\ket{\zeta_-}p_\nu\,p_\rho \EH_+^2 +3\, \langle a_{\mu}^\dagger \ket{\zeta_-}
\langle a_{\nu}^\dagger \ket{\zeta_+} p_\rho \EH_+\right)\right]
\ket{\hat\varphi(\t,p)}\0
\ee
\be
&&\left( \omega^{\mu\nu} \langle a_{\mu}^\dagger \ket{\zeta'_-}
\langle a_{\nu}^\dagger \ket{ \l_+} \ket{\hat\varphi(\t,p)}\right)*
\ket{\hat\Xi}+
\ket{\hat\Xi}*\left(\omega^{\mu\nu} \langle a_{\mu}^\dagger \ket{\zeta'_-}
\langle a_{\nu}^\dagger \ket{ \l_+} \ket{\hat\varphi(\t,p)}\right)
\label{thetalambdastar3}\\
&&= e^{-\frac 12 Gp^2}\omega^{\mu\nu}\left[ \frac 12\,
\langle a_{\mu}^\dagger \ket{\zeta'_-}\,
\langle a_{\nu}^\dagger \ket{ \l_+}+ \frac 12\,
\langle a_{\mu}^\dagger \ket{\zeta'_+}\,
\langle a_{\nu}^\dagger \ket{ \l_-}\right.\0\\
&&\left.\quad\quad\quad\quad\quad\quad
+\langle a_{\mu}^\dagger \ket{ \zeta'_-} p_\nu\, \bra{\t_0}\frac T{1-T^2}
\ket{\l_+} + p_\mu\,\langle a_{\nu}^\dagger \ket{ \l_-} \,\EH_+
\right]\ket{\hat\varphi(\t,p)} \0
\ee
and
\be
&&\left(g^{\mu} \langle a_{\mu}^\dagger\ket{r_-}
\ket{\hat\varphi(\t,p)}\right)*
\ket{\hat\Xi}+\ket{\hat\Xi}*
\left(g^{\mu} \langle a_{\mu}^\dagger\ket{r_-}
\ket{\hat\varphi(\t,p)}\right)\,\label{starr3}\\
&&\quad\quad\quad\quad\quad\quad=\,e^{-\frac 12 Gp^2}\,
g^{\mu} \langle a_{\mu}^\dagger\ket{r_-}
\ket{\hat\varphi(\t,p)}\0
\ee
where $\ket{ \l_-} =\lim_{\eta\to 0^+}  \ket{\eta_-}\l(\eta)$,
and $\ket{\zeta_+}= \lim_{\eta\to 0^+} \ket{\eta_+}\zeta(\eta)$,
$ \l(\eta)$ and $\zeta(\eta)$ being the same functions as above,
(\ref{expan'}).

Now, in order for the LEOM to be satisfied the sum of these
three terms, (\ref{thetastar3},
\ref{thetalambdastar3}) and (\ref{starr3}), must reproduce (\ref{level3st}).
From the second term in the RHS of (\ref{thetalambdastar3}) we see
that we must have $e^{-\frac 12 Gp^2}=4$, i.e. $p^2=-2$, the mass--shell
condition for level 3 states. This implies that, the RHS of the second equation
$\omega^{\mu\nu}\langle a_{\mu}^\dagger \ket{\zeta'_-}\,
\langle a_{\nu}^\dagger \ket{ \l_+}$ appears with a coefficient 2 in front,
therefore half of this term will reproduce (\ref{level3st}) and the other
half must be canceled against the other terms. Similarly
in the RHS of (\ref{starr3}) the term
$g^\mu \langle a_{\mu}^\dagger\ket{r_-}$ appears with a coefficient 4.
So 1/4 of it will reproduce (\ref{level3st}) and 3/4 will have to be
canceled.

Next, as in the previous subsection, we count the degrees of divergence
for $\eta\to 0$ of the various terms in the above three
equations, which is $-2$ for the first and third terms of the
RHS of (\ref{thetastar3}) and 0 for the remaining ones;
it is 0 for the first two terms in the RHS of (\ref{thetalambdastar3})
and --2 for the other two; finally it is zero for the term in the
RHS of (\ref{starr3}). Now what we have to do is collecting all the
unwanted terms in the RHS and imposing that the sum of the coefficients
in front of them vanish. From what we just said, we can deduce that
we must have
\be
&&\omega^{\mu\nu}\sim \theta^{\mu\nu\rho} p_\rho\0\\
&&\theta_{\mu}{}^{\mu\rho}\sim \theta^{\mu\nu\rho}p_\mu p_\nu\sim
a\,\omega^{\mu\rho}p_\mu+b\, \omega^{\rho\mu}p_\mu\label{simeq}\\
&& \omega^{\mu\rho}p_\mu\sim g^\rho\0
\ee
for some constants $a$ and $b$.
These are very close to (\ref{Vir3a},\ref{Vir3b}). However it must be
proven that the arbitrary constants
we have at our disposal (i.e. $x,y,u,v,z$ and the coefficients of
$\zeta(\eta),\lambda(\eta)$ and $r(\eta)$) are sufficient to satisfy
all the conditions. This is an elementary algebraic problem.
The straightforward calculations are carried out in Appendix G
where it is shown that all the conditions are met.
So we can conclude that
\beq
\ket{\hat \varphi(g,\omega,\theta,\t,p)} * \ket{\hat \Xi} +
\ket{\hat \Xi} *\ket{\hat \varphi(g,\omega,\theta,\t,p)}=
\ket{\hat \varphi(g,\omega,\theta,\t,p)}\label{verlevel3}
\eeq

\newsection{Cohomology}

A solution to the LEOM is not automatically a solution fit to
represent a physical string state. The reason for this is the huge
gauge invariance which soaks all physical states in SFT. Any
solution to the LEOM is in fact defined up to \be \Q_0 \Lambda
\equiv \Q \Lambda + \Phi_0 * \Lambda - \Lambda *
\Phi_0\label{gaugetrans} \ee where $\Phi_0$ is our reference
classical solution (see section 3) and $\Lambda$ is any string
state of ghost number 0. Only string field solutions which cannot
take the form of (\ref{gaugetrans}) are significant solutions and
can represent physical states. Phrased another way, $\Q_0$ is
nilpotent, therefore it defines a cohomology problem: only
nontrivial cohomology classes are physically interesting.
Unfortunately a systematic approach to this problem is missing
(although some progress can be found in \cite{imbimbo}), the more
so for VSFT. Partial elaborations on the gauge freedom in VSFT can
be found in \cite{HK, ima}.  In this section we will not try a
systematic approach to the cohomology problem. Nevertheless it
turns out to be rather easy to figure out $\Lambda$ `counterterms'
that `almost trivialize' the solutions we have found in the
previous section, but actually do not kill them at all. This makes
us confident that what we have found in the previous sections
singles out nontrivial cohomology classes.

To simplify the problem as much as possible we will exclude all the $\Lambda$'s with
a nontrivial ghost content. If $\Lambda$ is a  matter state tensored with the ghost
identity, see \cite{HK, ima}, then the gauge transformation (\ref{gaugetrans}) for a
(pure matter) state $\phi$
 can be written simply through $\Lambda$'s matter part as follows:
\be
\delta \phi = \hat \Xi *_m \Lambda - \Lambda *_m \hat\Xi\label{gt}
\ee
where $\hat \Xi$ is the dressed sliver.  Our problem is now to find
matter states $\Lambda$ such that (\ref{gt}) gives some of the
solutions we found in the previous sections. Let us try the following
one (we set $e=1$ and drop the label $m$ in $*_m$ throughout this section)
\be
\ket{\Lambda(g,\zeta)} =
g^\mu \langle(1+C)\zeta\,a_\mu^\dagger\rangle \ket {\hat \varphi_t(\t,p)}
\label{gtv}
\ee
where $\ket {\hat \varphi_t(\t,p)}$ is the tachyon wavefunction. The gauge
transformation (\ref{gt}) becomes
\be
&&\ket{\hat \Xi} * \ket{ \Lambda(g,\zeta)}
- \ket{\Lambda(g,\zeta)} * \ket{\hat\Xi}
\0 \\
&& \,=\, e^{-\frac 12 G\,p^2}
\left\{g^\mu \langle a^\dagger(\rho_1-\rho_2) (1+C)\zeta\rangle\,
+ \frac 1 {\k+1}\, g^\mu \langle a_\mu^\dagger
\left( \ket{\xi}\bra{\xi} - \ket{C\xi}\bra{C\xi}\right)
\frac {1}{1-T}\ket{(1+C)\zeta} \right.
\0 \\
&&\quad\quad\left.
\,-\,2\b\, (p\cdot g)\left[\bra{\xi}\frac T{1-T^2}\ket{(1+C)\zeta}
- \kappa\, \langle\xi|\frac 1{1-T^2}|(1+C)\zeta\rangle \right] \right\}
|\hat \varphi_t(\t,p)\rangle
\label{coho1}
\ee
Now suppose that $\rho_2\zeta=\zeta$ and $\rho_1\zeta=0$. We get
\be
\lefteqn{\ket{\hat \Xi} *\ket{ \Lambda(g,\zeta)}
- \ket{\Lambda(g,\zeta)} * \ket{\hat\Xi}}
\label{coho2} \\
&& =\, e^{-\frac 12 G\,p^2}\left[ - g^\mu \langle a^\dagger(1-C)\zeta\rangle
- 2\,\b\,(p\cdot g)\, \bra{\xi}\frac {T-\k}{1-T^2}\ket{\zeta}\right]
|\hat \varphi_t(\t,p)\rangle \0
\ee
Comparing now this with eq.(\ref{vectoransatz}) we see that, if we
choose the $\zeta$'s in the two equations to be the same, we set
$g^\mu=d^\mu$ and suitably normalize $ \Lambda(g,\zeta)$, the gauge
transformation (\ref{coho2}) gives back just the vector state
eigenfunction (\ref{vectoransatz}), or, in other words, the latter
belongs to the trivial cohomology class.

Therefore, if $\zeta(k)$ is a regular function for $k\sim 0$
(henceforth let us refer to such a $\ket{\zeta}$ as {\it regular} or
{\it smooth} at $k=0$) , the
vector state we have constructed in section 5.2 is cohomologically
trivial. In order to get something nontrivial we have to probe the
string midpoint. Therefore let us try with $\ket{\zeta} \sim C\ket{\eta}$
(from now on let us refer to the latter as {\it singular} or
{\it concentrated at $k=0$}). It satisfies $\rho_2\zeta=\zeta$ and
$\rho_1\zeta=0$ and $\ket{(1+ C)\zeta}\sim \ket{\eta_+}$,
$\ket{(1- C)\zeta}\sim \ket{\eta_-}$ (see eqs.(\ref{epsipm})).
Therefore, in this case too, as long as the parameter $\eta$ remains
finite, the vector state is trivial. One may be tempted to conclude
that also in the limit $\eta\to 0$, such a conclusion persist and
therefore the vector wavefunction we have defined be always trivial.
But this would be a sloppy deduction. For in the process of taking the
limit $\eta\to 0$ there emerges the true nature of cohomology.

For a cohomological problem to be well defined it is not enough to have
a nilpotent operator, one must also define the set of objects which
such an operator acts upon, i.e. the space of cochains. In our case
a precise definitions of the cochain space has not been given so far, and
it is time to fill in this gap. It is clear that the issue here is the
distinction between the states that vanish and those that do not vanish
in the limit $\eta\to 0$. For instance, (see (\ref{v+},\ref{v-})),
$\ket{0_+}$ belongs to the former set (let us call it an {\it evanescent} state)
while $\ket{0_-}$ belongs to the latter. We define the space of nonzero cochains
as the space of states that are finite in the limit $\eta\to 0$,
while the zero cochain is represented by 0. All this is well--defined
and makes up a linear space and it is the only sensible choice to define
a cohomology in this context (see Appendix H for a discussion of this point).

With the previous definition let us return to the vector eigenfunction. Thanks to
the discussion following eqs.(\ref{coho1},\ref{coho2}), we see immediately that
if $\zeta$ in (\ref{vectoransatz}) is smooth near $k=0$,
then the corresponding wavefunction is a coboundary. If, on the other hand,
$\zeta \sim \ket{C\eta}$, i.e. is concentrated at $k=0$, then the state is
a nontrivial cocycle, because we cannot
figure out any non--evanescent $\Lambda$ which generate it via (\ref{gt}):
the only one that does the job is evanescent.
This same conclusion can be drawn for the level 2 and level 3 states we found above
(which were formulated directly in terms of vectors concentrated at $k=0$).
All these states are cohomologically nontrivial.

At this point we can discuss also the implication of the simplifying assumption we
introduced in section 7.2 and 7.3, i.e. that the dressing function $\xi(k)$ is
non-vanishing only from a certain finite negative point down to $-\infty$ in the
$k$-axis. This assumption induced remarkable simplifications in our
analysis, but that was the only reason why it was introduced: one can do without it.
Anyhow let us ask ourselves what would have happened had we
introduced this assumption in the vector case. In the case of $\zeta$ being
concentrated at $k=0$ the last two terms at the RHS of (\ref{vectorsol})
would vanish and we would not need to impose the transversality condition
(\ref{transvers}). If, on the other hand, $\zeta$ is smooth at $k=0$ then,
in order to satisfy the LEOM, we would have to impose the transversality condition
(\ref{transvers}) together with the additional condition (\ref{condvector}),
but in this case we would get a trivial solution.
This conclusion seems to be paradoxical only if we forget the relation between
cohomology and Virasoro conditions. In fact it is perfectly logical.
First of all we should remember that we have two ways of expressing
the physicality of a given state. Either we say that this state is a
nontrivial cocycle defined up to generic coboundaries (this is the
cohomological way of putting it), or we impose conditions on the parameters
of the state (polarizations) in such a way that its indeterminacy  (coboundaries)
get suppressed (and this is the gauge fixing way). Now, the above apparent
paradox means that the simplifying assumption, which seems to suppress the
transversality condition on the nontrivial cocycle (singular $\zeta$),
can be made up for by adding to the solution a trivial cocycle (regular $\zeta$).
In other words, the simplifying assumption corresponds to partially fixing
the gauge freedom. It can be seen that this is true also in the more
complicated cases of level 2 and level 3.

With this remarks we end our analysis of cohomology in VSFT. This problem
would deserve of course
a more thorough treatment, but we believe we have caught some of the
essential features of it.

\newsection{Proliferating solutions}

All the solutions to the LEOM considered so far depend on three
parameters: $e,\a,\b$. As will be seen below, $e$ has to be
set equal to 1, but the other two parameters are free.
We wish to show in this section that the solutions to the LEOM are
even more general than this. In fact we can prove that, if
$\ket{\hat\varphi(\t,p)}$ is the matter part of the tachyon solution to the
linearized equation of motion, i.e. a solution to (\ref{EOMlm}),
then any state of the form
\beq
(\langle a^\dagger_{\mu_1} \xi_\pm\rangle\ldots\langle a^\dagger_{\mu_s}
\xi_\pm\rangle )
\ket{\hat\varphi(\t,p)}\label{gentach}
\eeq
where $\xi_\pm=(1\pm C)\xi$,
is also a solution for any $s$, with the same mass as the tachyon
for any random choice of the $\pm$ signs.
For
\be
\lefteqn{(\langle a^\dagger_\mu \xi\rangle\ket{\hat\varphi(\t,p)})
* \ket{\hat \Xi} + \ket{\hat \Xi} *
(\langle a^\dagger_\mu \xi\rangle\ket{\hat\varphi(\t,p)})}
\0 \\
&& = \langle a^\dagger_\mu \xi\rangle\left[\ket{\hat\varphi(\t,p)}
* \ket{\hat \Xi} + \ket{\hat \Xi}*\ket{\hat\varphi(\t,p)}\right]
= \langle a^\dagger_\mu \xi\rangle\ket{\hat\varphi(\t,p)}
\label{versoltach}
\ee

The derivation of the first equality is given in Appendix I.
The same can be shown if we replace $\langle a^\dagger_\mu \xi\rangle$ with
$\langle a^\dagger_\mu C\xi\rangle$.
This proves the above claim for $s=1$. But it is evident that now we can
proceed recursively by replacing in (\ref{versoltach})
$\ket{\hat\varphi(\t,p)}$ with
$\langle a^\dagger_\nu \xi\rangle\ket{\hat\varphi(\t,p)}$ and
$\langle a^\dagger_\nu C\xi\rangle\ket{\hat\varphi(\t,p)}$,
respectively, which
are also solutions, thereby proving the statement for $s=2$, and so on.

We refer to all these states as the
descendants of $\ket{\hat\varphi(\t,p)}$, or {\it tachyon descendants}.
We can easily define a generating state for them
\beq
\ket{\hat\varphi(g,\t,p)} =
e^{(g_+^\mu \langle a^\dagger_\mu \xi_+\rangle+g_-^\mu \langle
a^\dagger_\mu \xi_-\rangle)}
\ket{\hat\varphi(\t,p)}\label{gentachyon}
\eeq
By differentiating with respect to $g_\pm^\mu$ we can generate all the
solutions of the type $\ref{gentach}$.

A similar result holds also for the other (tensor) solutions
of the LEOM. At level $n$ such states take the form
\beq
\ket{\hat \varphi(\theta,n,\t,p)}\equiv\ket{\hat \varphi(\theta_1,...,
\theta_n,\t,p)} =
\sum_{i=1}^n
\theta^{\mu_1\ldots\mu_i}_i \langle a_{\mu_1}^\dagger \zeta^{(i)}_1\rangle
\ldots \langle  a_{\mu_i}^\dagger \zeta^{(i)}_i\rangle \ket{\hat\varphi(\t,p)}
\label{levelnst'}
\eeq
where the polarizations $\theta_i$ must satisfy constraints similar
to those found for level 1,2 and 3. As shown in Appendix I, we have a
result similar to the above. The LEOM is satisfied with the same mass
\be
&&(h^\nu\langle a^\dagger_\nu \xi\rangle\ket{\hat\varphi(\theta,n,\t,p)})*
\ket{\hat \Xi} + \ket{\hat \Xi}*
(h^\nu\langle a^\dagger_\nu \xi\rangle\ket{\hat\varphi(\theta,n,\t,p)})\0\\
&&=  h^\nu\langle a^\dagger_\nu \xi\rangle
\left[\ket{\hat\varphi(\theta,n,\t,p)}*\ket{\hat \Xi}
+\ket{\hat \Xi}*\ket{n,\hat\varphi(\theta,n,\t,p)}\right]
=  h^\nu \langle a^\dagger_\nu \xi\rangle
\ket{\hat\varphi(\theta,n,\t,p)}\label{versoln}
\ee
but, now, under some conditions: either
\beq
\bra{\xi} \frac {T-\kappa}{1-T^2}\ket{C\zeta^{(i)}_j}=0\label{vanishcond}
\eeq
(which is the case for instance when $\rho_2\zeta^{(i)}_j\!=\!\zeta^{(i)}_j,
\,\rho_1\zeta^{(i)}_j\!=\!0)$
or, if this is not true (as is the case in our previous analysis), the
polarization $h$ is transverse to the $\theta_i$'s
when contracted with the index $\mu_j$:
\beq
h^\nu\,\eta_{\nu\mu_j}\, \theta_i^{\mu_1\ldots\mu_j\ldots\mu_i}=0,
\label{transvn}
\eeq
and this must hold $\forall i,j,\quad 1\leq j\leq i ,\quad 1\leq i\leq n$.

Also here we can replace $\langle a^\dagger_\nu \xi\rangle$ with
$\langle a^\dagger_\nu C\xi\rangle$ and obtain a new solution with the
same mass, and therefore we can define the $\pm$ combination, as above.
Inductively we can prove that
\beq
(h_1^{\nu_1} \langle a^\dagger_{\nu_1} \xi\rangle
\ldots h_s^{\nu_s}\langle a^\dagger_{\nu_s} \xi\rangle)
\ket{\hat\varphi(\theta,n,\t,p)}
\label{statens}
\eeq
satisfy the LEOM with the same mass provided each $h_j$ is transverse to
each $\theta_i$ on all indices. We can then introduce $C$ in every
$\langle a^\dagger\ket{\xi}$ factor and obtain new independent solutions.
It is evident that the most general state with the same mass takes the
form
\beq
\langle a_{\nu_1}^\dagger\xi_\pm\rangle \ldots
\langle a_{\nu_s}^\dagger\xi_\pm\rangle
\sum_{i=1}^n
\theta^{\nu_1\ldots\nu_s;\mu_1\ldots\mu_i}_i
\langle a_{\mu_1}^\dagger \zeta^{(i)}_1\rangle
\ldots \langle a_{\mu_i}^\dagger \zeta^{(i)}_i\rangle \ket{\hat\varphi(\t,p)}
\label{mostgen}
\eeq
with generic $s$, provided the tensor $\theta_i$ are traceless when
any index $\nu$ is contracted with any index $\mu$.
However, any state of the type (\ref{mostgen}) is a finite linear
combination of states of type (\ref{levelnst'}). A generating function
for the latter is
\beq
\ket{\hat\varphi(h,\t,p)} =
e^{(h_+^\mu \langle a^\dagger_\mu \xi_+
\rangle+h_-^\mu \langle a^\dagger_\mu \xi_-\rangle)}
\ket{\hat\varphi(\theta,n,\t,p)}\label{genfn}
\eeq
Differentiating with respect to $h_\pm$ the required number of times,
we can construct any state of the type (\ref{levelnst'}). A generating
function is particularly useful in computing norms or amplitudes.

To finish this section a comment is in order concerning the
enormous proliferation of solutions to the linearized equations
of motion. All the states we have found seem to be cohomologically
nontrivial on the basis of the analysis in the previous section.
The existence of an infinite tower of descendants of
a given solution is, generically speaking, hardly a surprise. We notice
that a similar
phenomenon is familiar in field theory. If $\phi_0(x)$ is a solution
to the Klein--Gordon equation $(\partial^2 +m^2)\phi=0$, then all the
derivatives of $\phi_0$ are solutions with the same mass.
We conjecture that here we are coming across something similar,
although the difference among different states of each tower
is given here not by the application of the space derivatives
(i.e. by powers of $\hat p$), but rather by the application of the
creation operators $a^\dagger_n$, $n>0$.

But now, the important question is: what is the nature of these
states? They seem to be physical, so it is important to clarify whether
they are simple copies of the first state of the tower (the {\it parent} state,
not containing $\langle a^\dagger \xi\rangle$ factors in their $\CP$ polynomial)
or have a different physical meaning. Looking at the generating state
(\ref{gentachyon}) one can see that, if $g_\pm\sim p$, this turns into
a redefinition
of the arbitrary constants $\a$ and $\b$ (see section 4.1). Therefore,
since these constants do not enter into physical quantities, such as $G$,
(they might appear in quantities like $H$, see below,
which is not by itself physical) we conclude that
the states of this type are copies of the tachyon
eigenfunction, without any physical differentiation from it.
It is possible to see
that this is true for any other tower of solutions. So the proliferation
we find seems to be a proliferation of representatives of physical states
(much in the same way as in the Coulomb representation of CFT we have two
representatives for any vertex). This redundancy of representatives,
which, it should be stressed, is due to dressing, may be a residue of the
gauge symmetry of VSFT.

\newsection{On the D25--brane tension}
\label{sec:states}

One of the unsatisfying aspects of the sliver in operator formalism
was the disagreement between the energy density of the classical
solution and the brane tension computed via the 3-tachyon
on-shell coupling. In this section we would like to show that
our approach can lead to a solution of this problem.

\subsection{3-tachyon on-shell coupling}
\label{ssec:3tach}

The cubic term of the VSFT action evaluated for 3 on-shell
tachyon fields should be equal to $g_T/3$, where $g_T$ is the
3-tachyon coupling constant for the open string, i.e.,
\begin{equation}
g_T \,=\, \frac{1}{g_0^2} \, \langle \, \varphi_t(\t,p_1) \,
| \, \varphi_t(\t,p_2) * \varphi_t(\t,p_3) \, \rangle \,
\Big |_{p_1^2=p_2^2=p_3^2=-m_t^2=1}
\end{equation}
Here $|\varphi_t(\t,p)\rangle$ must be normalized so as to give the
canonical
kinetic term in the low-energy action (see \cite{HKw}, Sec.\ 5.2).
Using (\ref{etachyon}), an explicit calculation gives
\begin{equation} \label{gt2}
g_T^2 \,=\, \frac{8g_0^2}{G^3} \, A^{13} \, \tilde{A}^{-1} \, \exp(-6H)
\end{equation}
where (see Appendix \ref{app:Hcalc})
\begin{equation} \label{hdress}
H \,=\, H_0 - \frac{(f_e-1)^2(\k+f_e)^2}{2(f_e+1)(f_e^3-1)}
\left[ \left(\frac{1}{\k+f_e}-\a\right)^2
\langle\t_0|\frac 1{1+T}|\xi\rangle^2 - \b^2 \right]
\end{equation}
and
\begin{equation} \label{agt}
A \,=\, \frac{[\det(1-\hat{T}_{e_1}\hat{T}_{e_2})]^3}{
[\det(1-\hat{\mathcal{T}}_{e_1e_2e_3}\mathcal{M}_3)]^2}
\,=\, \frac{(f_1f_2-1)^6}{(f_1f_2f_3-1)^4}
\frac{\left[\det(1-T^2)\right]^3}{\left[
\det(1-T\mathcal{M}_3)\right]^2}
\end{equation}
$\tilde{A}$ is obtained from $A$ by replacing all the relevant objects
with tilded ones (ghost part).
$H_0$ is a naively vanishing `bare' term. However
in level truncation it turns out to be nonvanishing
due to the so--called `twist anomaly' \cite{HM1, HM2}.

It was shown by Okuyama that the ratio of determinants in
the RHS of (\ref{agt}) diverges like $L^{5/18}$ as $L\to\infty$.
Similarly, the corresponding term in
$\tilde{A}$ behaves as $L^{11/18}$. Now, in order for $g_T$ to be finite,
the only possibility is to tune the ``dressing'' parameter
$e$ to the value 1 in some suitable way. This is the reason why,
as anticipated many times in the previous sections, we have to set $e=1$.
But in the formula (\ref{agt}) this has to be done with an appropriate
scaling of $e$ to 1, in such a way as to get an overall finite result.
This is very close to what we did in \cite{BMP1} to make the
dressed sliver action finite. Following the same prescription,
we render separately finite $A$ and $\tilde{A}$
(the matter and ghost part). This entails that $H$ must be finite too.
It is easy to see that the only way to implement this is to
let $f_e\to1 $ (i.e. $e\to 1$) in such a way that
\begin{equation}
f_e-1 = s_t L^{-5/36} \qquad\mbox{and}\qquad
f_{\tilde{e}} - 1 = \tilde{s}_t L^{-11/36}
\end{equation}
where $s_t$ and $\tilde{s}_t$ are constants. We note that $f_e$ and
$f_{\tilde{e}}$ scale the same way as $f_\epsilon$ and
$f_{\tilde{\epsilon}}$ in \cite{BMP1}.

Using $f_e\to1$ in (\ref{hdress}) we obtain $H=H_0$. From (\ref{gt2})
it then follows that $g_T$ is independent of the dressing parameters
$\a$ and $\b$. We expect this to be true for all physical quantities.

As in the case of the energy of the dressed sliver, the precise value
of $g_T$ depends not only on the value of the (so far undetermined)
scaling parameter $s_t$, but also on the way in which the multiple
limit $f_1,f_2,f_3\to1$ is taken. Now we would like to argue that,
with the proper choice of limit prescriptions, two problems, which affect
the approach with the standard sliver, may be solved:
\begin{itemize}
\item
Validity of EOM and LEOM when contracted with the solutions themselves.
\item
Correct value of the product of the sliver energy density and $g_T^2$.
\end{itemize}

\subsection{Scaling limit}
In general observables contain such terms as $(f_1f_2-1)$ and/or
$(f_1f_2f_3-1)$.
In the scaling limit $f_i-1\approx s_iL^x$, where $x<0$ and
$L\to\infty$, one expects
\begin{equation}
(f_1f_2-1) \,\approx\, s_{12} L^x \;, \qquad
(f_1f_2f_3-1) \,\approx\, s_{123} L^x
\end{equation}
but the scaling coefficients $s_{12}$ and $s_{123}$ are a priori not
unique. They depend on the precise prescription for taking the multiple
limits (see \cite{BMP1}, Sec.\ 5 and Appendix C).

In Sec.\ 5 of \cite{BMP1} it was shown that there is a connection
between the prescription for taking limits and the validity of the EOM.
Considering the EOM for the
dressed sliver contracted with the dressed sliver, we have
\begin{eqnarray} \label{dskin}
&& \langle\, \hat{\Xi}_{\epsilon_1\tilde{\epsilon}_1} | \mathcal{Q}
\,|\, \hat{\Xi}_{\epsilon_2\tilde{\epsilon}_2} \rangle
\,=\, \left(1-\frac{1}{f_1f_2}\right)^{\!-26}
\left(1-\frac{1}{\tilde{f}_1\tilde{f}_2}\right)^{\!2}
\langle\,\Xi|\mathcal{Q}|\,\Xi\rangle
\\ &&
\langle\, \hat{\Xi}_{\epsilon_1\tilde{\epsilon}_1} |\,
\hat{\Xi}_{\epsilon_2\tilde{\epsilon}_2} *
\hat{\Xi}_{\epsilon_3\tilde{\epsilon}_3} \rangle
\,=\, \left(1-\frac{1}{f_1f_2f_3}\right)^{\!-26}
\left(1-\frac{1}{\tilde{f}_1\tilde{f}_2\tilde{f}_3}\right)^{\!2}
\langle\,\Xi |\, \Xi * \Xi\rangle \qquad
\end{eqnarray}
where $|\Xi\rangle=|\hat{\Xi}_0\rangle$ is Hata and Kawano's sliver. Let us
denote
\begin{equation}
\zeta_{cc} \,=\, - \frac{\langle\,\Xi|\mathcal{Q}\,|\,\Xi\rangle}{
\langle\,\Xi |\, \Xi * \Xi\rangle}
\end{equation}
If the EOM holds for this sliver solution one gets $\zeta_{cc}=1$.
However, it was argued
in \cite{HM3} that this may not be the case in the level truncation
regularization. We believe that this `anomaly' should be resolved within
the level truncation scheme and we expect (see below) that the result
should be $\zeta_{cc}=1$. However we would like to point out that
the formalism we have presented in this paper can allow also for
values of $\zeta_{cc} \neq 1$. So, to keep this possibility into account,
we leave $\zeta_{cc}$ generic. In fact, as we will see,
this variable can be absorbed by the dressing.

From the requirement that `contracted' EOM be satisfied
\begin{equation}
\lim_{\epsilon_i,\tilde{\epsilon}_j\to1}
\langle\, \hat{\Xi}_{\epsilon_1\tilde{\epsilon}_1} | \mathcal{Q}
\,|\, \hat{\Xi}_{\epsilon_2\tilde{\epsilon}_2} \rangle
\,=\, - \lim_{\epsilon_i,\tilde{\epsilon}_j\to1}
\langle\, \hat{\Xi}_{\epsilon_1\tilde{\epsilon}_1} |\,
\hat{\Xi}_{\epsilon_2\tilde{\epsilon}_2} *
\hat{\Xi}_{\epsilon_3\tilde{\epsilon}_3} \rangle
\end{equation}
we obtain the following condition on the scaling parameters
\begin{equation}
\left( \frac{s_{ccc}}{s_{cc}} \right)^{\!-26}
\left( \frac{\tilde{s}_{ccc}}{\tilde{s}_{cc}} \right)^{\!2}
\,=\, \zeta_{cc} \;.
\end{equation}
We see that a possible anomaly in the contracted EOM can be
cured by an appropriate limit
prescription. However it should be noticed that the limit
prescription to be used in such a case is
not a priori clear and far from simply describable. We
recall that in \cite{BMP1} (see section 2)
we defined a privileged way of taking this kind of limits:
the nested limits prescription. In the
light of the analysis of \cite{BMP1} this looked as the most
natural prescription. Any other way
seems to be artificial. This is the reason why we tend to
believe that there should not be any
anomalous $\zeta_{cc}$.

In the case of the `contracted' LEOM for our tachyon solution
(\ref{tachim})
\begin{equation}
\langle \hat\phi_e(\t,p) | \mathcal{Q}_0|\hat \phi_e(\t,p) \rangle \,=\, 0
\end{equation}
the possible anomaly \cite{HM3,RSZ4} is cured by taking
\begin{equation}
\zeta_{tt}\equiv -\frac{\bra{\phi_t}\Q \ket{\phi_t}}{2\bra{\phi_t}
\phi_t*\Xi\rangle}=
 \left( \frac{s_{ttc}}{s_{tt}} \right)^{\!-26}
\left( \frac{\tilde{s}_{ttc}}{\tilde{s}_{tt}} \right)^{\!2}\label{zetatt}
\end{equation}
where $\phi_t$ is the undressed tachyon $e=0$
(from the symmetry of 3-string vertex for cyclic permutations it follows
$s_{ttc}=s_{tct}=s_{ctt}$).


\subsection{D25-brane energy}
Let us now calculate the product of the dressed sliver energy
density and $g_T^2$, which if our dressed sliver represents the D25-brane,
should be
\begin{equation} \label{ecgt2st}
\left(E_c \, g_T^2\right)_{OST} \,=\, \frac{1}{2\pi^2}
\end{equation}
From (\ref{dskin}) and (\ref{gt2}) we obtain
\begin{equation} \label{dsecgt2}
E_c \, g_T^2 \,=\, \left( \frac{s_{tt}}{s_{cc}} \right)^{\!26}
\left( \frac{\tilde{s}_{tt}}{\tilde{s}_{cc}} \right)^{\!-2}
\left( \frac{s_{ttt}}{s_{tt}} \right)^{\!-52}
\left( \frac{\tilde{s}_{ttt}}{\tilde{s}_{tt}} \right)^{\!4}
\left(E_c \, g_T^2\right)_0
\end{equation}
where $(E_c \, g_T^2)_0$ is the result for the standard
sliver\footnote{ It should be mentioned that the calculation of
$g_T$ using the definite--twist tachyon (\ref{tachim}) shows a
dependence on the parameter $\beta$. This dependence may of course
be absorbed within the scaling parameters.}. In \cite{HM2,RSZ4} it
was shown that $(E_c \, g_T^2)_0$ is given by
\begin{equation}
(E_c \, g_T^2)_0 \,=\, \frac{\pi^2}{3}\left(\frac{16}{27\ln2}\right)^3
\end{equation}
which is obviously different from (\ref{ecgt2st}).

Note that scaling parameters $s_{ttt}$ and $\tilde{s}_{ttt}$ do not
appear in any LEOM and so are not affected by the analysis of the
previous subsection. Therefore they can take values such that
(\ref{ecgt2st}) is satisfied for the dressed sliver.

The possibility we have just pointed out is important because it
removes a sort of no--go theorem, \cite{HM3},
that seemed to exist in the operator treatment of the sliver
solution. However we should point out
that there is a difference between the limiting/tuning procedure
used in \cite{BMP1} to define a
finite energy density of the dressed sliver and the same procedure
used here in order to obtain the
matching between RHS and LHS of (\ref{ecgt2st}). In the first
case the critical dimension was behind
the argument we used and supported it (see Appendix K), in the latter case we have
not been able to find a similar
argument in favor of our tuning procedure. Without this the theory
has apparently lost some of the predictability: see, for instance,
(\ref{dsecgt2}) which
is undetermined without knowing $s_{ttt}$ and $\tilde{s}_{ttt}$.
However we believe that such an argument should exist which relates
 tuning to the consistency of the whole theory (of which we have explored
only a minute part).

\newsection{Discussion}

In this paper we have addressed the problem of finding the
open string spectrum on a D25--brane in
the operator formulation of VSFT. From previous works,
\cite{HKw, HM3, HK, Hgauge}, it looked like
the (standard) sliver solution is unable to capture
such a spectrum with the expected
physicality conditions (Virasoro constraints); in particular
it was not possible to derive the transversality condition for
the $U(1)$--gauge field and to describe lower spin components
of massive excitations.

To start with in this paper we have proved that the dressing creates
extra structure whereby the photon transversality can be accounted for.
Then we have coped with the problem of higher level states, where we have
come to terms with the crucial role played by the $k\sim 0$ region.
Our analysis implies that one has to take into
account a twist--even $k=0$ mode of the $k$--spectrum.
This mode is usually disregarded because
it vanishes when contracted with any Fock space operator.
Due to the star product, however, it becomes essential
for the consistency of the formulas necessary in our calculations,
in particular it preserves the validity of many of the properties
of the 3--strings vertex Neumann coefficients when
the $k\sim 0$ region is probed.

We have also analyzed part of the cohomology problem implicit in the LEOM
which generates on--shell fluctuations of the background
D25--brane. In view of our analysis we find that two sets
of modes are not gauge trivial. The first
set covers the whole open string spectrum (our analysis
stopped at level 3, i.e $m^2=2$, but we do not see any conceivable
obstruction in going further). The
second set is `orthogonal' to the first and consists
of states constructed by applying powers of
$\xi_n a^\dagger_n$ factors to the states of the first set.
They give rise to an infinite tower of
descendants for every physical state, but they seem to describe the same
observables as their
parent states, thus creating a degeneracy of `representatives'.
A possibility is that this (numerable) redundancy is
inherited by the gauge invariance of SFT and, thus,
that it may be gauge fixed via some more refined study of the
gauge structure of VSFT that takes into account the ghost
sector.\footnote{Some analysis of
this kind was performed in \cite{Hgauge} in order to implement the
$U(1)$ gauge transformations on
the massless vector.}

Finally we turned our attention to  the well known problem of
matching  the energy
density of the classical solution with the D25--brane tension
computed via the 3 on--shell tachyons
coupling; although this problem was resolved in \cite{Oka} in
the BCFT formulation of VSFT,  it
remained an open puzzle for the operator formalism \cite{HM3}.
By extending the analysis started in
\cite{BMP1},  we showed that the arbitrariness we have in tuning
the dressing parameters $e$, to
the level truncation cutoff, can be used  to satisfy the LEOM,
when contracted with the solutions
themselves.  Appropriately choosing the scaling limit, we can obtain the
correct value for the energy density expressed through the 3--tachyons
coupling constant. However, at present, this value can only be
accommodated, not uniquely determined.

Due to the extensive matter dealt with in this paper it was
inevitable to leave aside or only partially treat a number of
issues. We have already pointed out that our analysis of the
dressed sliver spectrum lacks of an algorithm to find general
solutions to the LEOM. Our cohomological analysis of the spectrum
is also incomplete. Finally a comment about regularizations is in
order. It looks like we are using three different regularizations
of the $k$--spectrum: the level truncation to evaluate $G_0$ and
$H_0$, Okuyama's prescription to evaluate determinants and the
$\eta$--regulator to compute LEOM's solutions. As we have shown in section 6 the $\eta$--regularization is based on the
density of eigenvalues formula (6.1) in the $L\to \infty$ limit.
Okuyama's regularization, \cite{Oku2}, and the twist anomaly computation of \cite{HM2}
are also based based on the same formula and use $1/\log L$ as a
regulator.
In section 6 we have shown that, for large $L$, the identification
(6.12) is justified. Therefore the three regularization procedures are
based on the same regulator and they must lead to the same
regularized quantities. A direct comparison of the results obtained with
the three different procedures is, however, in general, not easy because they
are devised to compute different objects in different contexts ($G_0$ and
$H_0$ with the large $L$ level truncation, infinite matrix determinants with
Okuyama's prescription and state polarizations with the $\eta$ regulator).
It would certainly be desirable to have a unified prescription.

Even with these cautionary remarks, we believe
we have produced strong evidence that, if  VSFT is properly
regularized, it can consistently describe the physical content
(both perturbative and non--perturbative) of bosonic string theory.
It may therefore be a useful tool in tackling more
challenging problems like the search for time--dependent solutions,
\cite{Hroll} and open--closed string duality, \cite{Senroll}.

\begin{center}
{\bf Acknowledgments}
\end{center}
We would like to thank Camillo Imbimbo for discussions.
C.M. would like to thank Theoretical Physics Department of University
of Zagreb for their kind hospitality during part of this research.
P.P. would like to thank SISSA--ISAS (Trieste) and ICTP (Trieste) for
their kind hospitality.
This research was supported by the Italian MIUR under the program
``Teoria dei Campi, Superstringhe e Gravit\`a'', and by Croatian
Ministry of Science, Education and Sports under the contract No.\
0119261.

\section*{Appendices}
\appendix

\newsection{A collection of useful formulas}

In this Appendix we collect some useful results and formulas involving the
matrices of the three strings vertex coefficients.

To start with, we recall that
\begin{itemize}
\item (i) $V_{nm}^{rs}$ are symmetric under simultaneous exchange of
the two couples of indices;
\item (ii) they are endowed with the property of cyclicity in the
$r,s$ indices, i.e. $V^{rs}= V^{r+1,s+1}$, where $r,s=4$ is
identified with $r,s=1$.
\end{itemize}

Next, using the twist matrix $C$  ($C_{mn}= (-1)^m \delta_{mn}$), we define
\beq
X^{rs} \equiv C V^{rs}, \quad\quad r,s=1,2,\label{EX}
\eeq
These matrices are often rewritten in the following way $X^{11}=X,\,
X^{12}=X_+,\, X^{21}=X_-$. They commute with one another
\beq
[X^{rs}, X^{r's'}] =0, \label{commute}
\eeq
moreover
\beq
CV^{rs}= V^{sr}C ,\quad\quad CX^{rs}= X^{sr}C
\eeq
Next we quote some useful identities:
\be
&&X^{11}+ X^{12}+ X^{21} = 1\0\\
&& X^{12}X^{21} = (X^{11})^2-X\0\\
&& (X^{12})^2+ (X^{21})^2= 1- (X^{11})^2\0\\
&& (X^{12})^3+ (X^{21})^3 = 2 (X^{11})^3 -  3(X^{11})^2 +1
\label{Xpower}
\ee
and
\beq
\frac{1-TX}{1-X}=\frac 1{1-T},\quad\quad \frac{X}{1-X}=\frac T{(1-T)^2}
\label{TXid}
\eeq
Using these one can show, for instance, that
\be
 \K^{-1} &=& \frac 1{(1+T)(1-X)} \left(\matrix {1-TX & TX_+\cr TX_- &1-TX\cr}
\right)\0\\
&=&\frac 1{1-T^2} \left(\matrix {1 & T(\rho_1-T\rho_2)\cr T(\rho_2-T\rho_1) &1\cr}
\right)\0\\
\M\K^{-1} &=&\frac 1{(1+T)(1-X)} \left(\matrix {(1-T)X  &X_+ \cr
X_-&(1-T)X\cr}\right)\0\\
&=& \frac 1{1-T^2} \left(\matrix {T & \rho_1-T\rho_2\cr \rho_2-T\rho_1 &T\cr}
\right)\label{id1}
\ee

Where we have defined the left/right Fock space projectors
\be
\rho_1 \!&=&\! \frac 1{(1 +T)(1-X)} \left[ X^{12} (1-TX)
+T (X^{21})^2\right]\label{rho1}\\
\rho_2 \!&=&\! \frac 1{(1 +T)(1-X)} \left[ X^{21} (1-TX)
+T (X^{12})^2\right]\label{rho2}
\ee
They satisfy
\beq
\rho_1^2 = \rho_1,\quad\quad \rho_2^2 = \rho_2, \quad\quad
\rho_1+\rho_2 = 1\quad\quad \rho_1\rho_2 = 0\label{proj12}
\eeq
i.e. they project onto orthogonal subspaces. Moreover,
\beq
\rho_1^T=\rho_1 = C\rho_2 C,\quad\quad
\rho_2^T=\rho_2 = C\rho_1 C.\label{rhorels}
\eeq
where $^T$ represents matrix transposition. As was shown in
\cite{RSZ3}, $\rho_1,\rho_2$ project out half the string modes.
Using these projectors one can prove that
\be
(X_+,X_-)\,\K^{-1} = (\rho_1,\rho_2),\quad\quad
\M \K^{-1} \T \left(\matrix {X_-\cr X_+\cr}\right) =
\left(\matrix {TX\rho_2 +TX_+\rho_1\cr TX_-\rho_2+TX\rho_1\cr}\right)
\label{id2}
\ee
which are used throughout the paper.

The following relations are often useful
\beq
\rho_1X_+ +\rho_2X_- = 1-XT,\quad\quad \rho_1X_- +\rho_2X_+ =X(T-1)
\label{Xrho}
\eeq

The next set of equations involve $\v_0,\v_\pm$.
We start with
\be
&&\v_++\v_- +\v_0 =0\0\\
&&\v_0^2+\v_+^2+\v_-^2=\frac 43 V_{00}\label{v0v+v-}\\
&&\v_0\v_- +\v_0\v_++\v_-\v_+= - \frac 23 V_{00}\0
\ee
Next we have the representation in terms of $\v_0$
\be
&&\v_+ = \frac 1{1+T}\left[(T-2)\rho_2 +(1-2T)\rho_1\right]\v_0\0\\
&&\v_- =\frac 1{1+T}\left[(T-2)\rho_1 +(1-2T)\rho_2\right]\v_0\0
\ee
from which we  get
\be
&&\v_+-\v_0 = -\frac 3{1+T} (\rho_2+T\rho_1)\v_0\0\\
&&\v_+-\v_- = -3 \frac{1-T}{1+T} (\rho_2-\rho_1)\v_0\label{v-v}\\
&&\v_--\v_0 = -\frac 3{1+T} (\rho_1+T\rho_2)\v_0\0
\ee
Using these equations in (\ref{v0v+v-}) it is easy to obtain in particular
\beq
\frac 23 V_{00} = 3 \langle \v_0 |\frac {T^2-T+1}{(1+T)^2}|\v_0\rangle=
\langle \t_0 |\frac 1{1+T}|\v_0\rangle\label{V00}
\eeq
where $\t_0= 3\frac{T^2-T+1}{T+1}|\v_0\rangle$.

In this work we use the continuous basis to evaluate various brackets
which appear in the computations. We therefore need
the matrices and vectors that define the 3 strings vertex in the $k$--basis.
We use normalized $k$-vectors, see \cite{Oku3},
\be
\ket{k}=\sum_{n=1}^{\infty} \frac 1k \sqrt{\frac{nk}{2\sinh\frac{\pi k}{2}}}
\oint\frac{dz}{2\pi i} \frac{1}{z^{n+1}}\left(1-\exp(-k\tan^{-1}z)\right)\ket n\0
\ee
so that
\be
\bra k k'\rangle=\delta(k-k')\0
\ee

With these conventions we have
\be
X = \int_{-\infty}^{\infty} dk X(k) \ket k \bra k,&\!\quad\!&
X(k)=-\frac{1}{1+2\cosh{\frac{\pi k}{2}}}\0\\
T = \int_{-\infty}^{\infty}dk T(k) \ket k \bra k, &\!\quad\!&
T(k) = - {\rm e}^{-\frac{\pi|k|}{2}} \0\\
\rho_1=\int_0^\infty dk\,\ket k \bra k, &\!\quad\!&
\rho_2=\int_{-\infty}^0 dk\,\ket k \bra k\label{matcont}
\ee
and
\be
\ket{\v_0} = \int_{-\infty}^{\infty}dk v_0(k) \ket k &\!\quad\!&
v_0(k) = - \frac{4}{3k} \sqrt{\frac{k}{\sinh\frac{\pi k}{2}}}
 \frac{\sinh^2\frac{\pi k}{4}}{1+2\cosh\frac{\pi k}{2}} \0\\
\ket{\t_0} = \int_{-\infty}^{\infty} dk t_0(k) \ket k &\!\quad\!&
t_0(k) = - \frac{4}{k\left(e^{\frac{\pi |k|}{2}}-1\right)}
 \sqrt{\frac{k}{\sinh\frac{\pi k}{2}}} \sinh^2\frac{\pi k}{4}
\label{vectcont}
\ee

All other matrices and vectors can be easily obtained using the
properties (\ref{Xrho}) and (\ref{v-v}). Notice that, since
$C\ket{k}=-\ket{-k}$, twist even vectors are represented by odd
functions and viceversa.

Notice also that $\t_0$ has a jump discontinuity in $k=0$
\be
t_0(0^+) = - t_0(0^-) = -\sqrt{\frac\pi 2} \0
\ee

\newsection{Solving for $\t_+$ and $\t_-$}

To solve for $\t=\t_++\t_-$ in the LEOM in full generality, we
reintroduce the parameter $\e$ in the equation of motion (\ref{EOMlm}).
This means deforming it as follows
\beq
\exp[- \t' a^\dagger \hat p]|\hat\Xi_{e*\e}\rangle
= |\hat\Xi_\e \rangle * (\exp[- \t a^\dagger \hat p]|\hat\Xi_e \rangle)
+ (\exp[- \t a^\dagger \hat p]|\hat\Xi_e \rangle) *|\hat\Xi_\e \rangle
\label{defEOMm}
\eeq
This seems to be a sensible deformation of (\ref{EOMlm}), since we know that,
as $\e \to 1$, $\hat\Xi_{e*\e}\to \hat\Xi_e$. As for $\t'$, this
deformation makes sense only if $\t' \to \t$ as $\e\to 1$.
This is indeed what happens.

In the following we will find a solution to (\ref{defEOMm}) and then take
the limit for $\e\to 1$.
\be
&&\t'_+= \v_0-\v_- +(X_+,X_-)\, \hat \K_{\e e}^{-1}\,\hat\T_{\e e}\,\left(
\matrix{\v_--\v_+\cr \v_+-\v_0}\right) + (X_+,X_-)\, \hat \K_{\e e}^{-1}\,
\left(\matrix{0\cr \t_+}\right)\label{t'+}\\
&&\t'_-=   (X_+,X_-)\, \hat \K_{\e e}^{-1}\,
\left(\matrix{0\cr \t_-}\right)\label{t'-}
\ee
We rewrite eq.(\ref{t'+}) in a more explicit form, using the methods
and results of Appendix B of \cite{BMP1}. In particular we need
the formula
\beq\label{1-P12mk}
(1-\CP_{\e e}\M\K^{-1})^{-1}\CP_{\e e} \,=\,
\frac{1}{B_{e\e}}\left(\matrix{
ef_e & \e(\rho_1-\kappa\rho_2) \cr e(\rho_2-\kappa\rho_1)
 &  \e f_\e\cr}\right)\CP_{\e e}
\eeq
where
\be
\CP_{\e e}=\left(\matrix{\e & 0 \cr 0& e}\right)\,P \;, \qquad
B_{e\e}=1+(1-e)(1-\e)\k \;. \0
\ee
Then eq.(\ref{t+}) can be rewritten as follows
\be
&&\t'_+= \v_0-\v_- +(X_+,X_-)\, \K^{-1}\,\T\,\left(
\matrix{\v_--\v_+\cr \v_+-\v_0}\right) + (X_+,X_-)\, \K^{-1}\,
\left(\matrix{0\cr \t_+}\right)\0\\
&&~~~~ + \frac 1{B_{\e e}} (\rho_1,\rho_2) \left(\matrix{e f_e
&\e(\rho_1-\k\rho_2) \cr e(\rho_2-\k\rho_1)& \e f_\e\cr}\right) \,
\CP_{\e e} \cdot\label{t+1}\\
&&~~~~~~~~\cdot \left[\left(\matrix{\frac 1{1-T^2}&\frac {TX_+}{(1+T)(1-X)}\cr
\frac {TX_-}{(1+T)(1-X)}&\frac 1{1-T^2}\cr}\right) \,
\left(\matrix{3\frac{1-T}{1+T}(\rho_2-\rho_1)|\v_0\rangle\cr
- \frac 3{1+T}(\rho_2+T\rho_1)|\v_0\rangle\cr}\right)\right.\0\\
&&~~~~~~~~~~~~
+ \left.\left(\matrix{\frac T{1-T^2}&\frac {X_+}{(1+T)(1-X)}\cr
\frac {X_-}{(1+T)(1-X)}&\frac T{1-T^2}\cr}\right)\,\left(\matrix{ 0\cr
\t_+\cr}\right)\right]\0
\ee
Carrying out the algebra one finds
\be
\t'_+=\rho_2 \t_+ + \rho_1\t_0 +\frac 1{\k+f_\e f_e} \!&&\! \left[
(1-f_\e)|\xi\rangle \, \langle\xi|\frac 1{1-T^2}|\t_0\rangle -
|C\xi\rangle \langle \xi|\frac {f_e+T}{1-T^2}|\t_0\rangle\right.\0\\
&& \left. + (f_\e-1)|\xi\rangle \, \langle\xi|\frac T{1-T^2}|\t_+\rangle
+ |C\xi\rangle \langle\xi|\frac {f_e+T}{1-T^2}|\t_+\rangle\right],\0
\ee
Applying now $C$ to both sides of this equation and summing the two
we get a C--symmetric equation.
\be\label{t++}
2\t'_+ = \t_+ + \t_0  + \!&&\! \frac{1}{\k+f_\e f_e} \left[
(1-f_\e)|\xi+C\xi\rangle \, \langle\xi|\frac 1{1-T^2}|\t_0\rangle -
|\xi+C\xi\rangle \langle C\xi|\frac{f_e+T}{1-T^2}|\t_0\rangle\right.\0\\
&&\left. + (f_\e-1)|\xi+C\xi\rangle \,
\langle\xi|\frac T{1-T^2}|\t_+\rangle + |\xi+C\xi\rangle
\langle\xi|\frac{f_e+T}{1-T^2}|\t_+\rangle\right]
\ee
Taking the difference we get instead
\be\label{t+-}
0 = (\rho_2-\rho_1) (\t_+ - \t_0) + \!&&\! \frac{1}{\k+f_\e f_e}
\left[(1-f_\e)|\xi-C\xi\rangle \, \langle\xi|\frac 1{1-T^2}|\t_0\rangle
+ |\xi-C\xi\rangle \langle C\xi|\frac {f_e+T}{1-T^2}|\t_0\rangle\right.\0\\
&&\left. + (f_\e-1) |\xi-C\xi\rangle \,
\langle\xi|\frac T{1-T^2}|\t_+\rangle - |\xi-C\xi\rangle
\langle\xi|\frac {f_e+T}{1-T^2}|\t_+\rangle\right]
\ee
Recalling that $(\rho_1-\rho_2)^2=1$, we multiply the last equation by
$\rho_1-\rho_2$ and obtain
\be\label{t+-'}
\t_+ = \t_0 -\frac{1}{\k+f_\e f_e} \!&&\! \left[ (1-f_\e) \,
\langle\xi|\frac 1{1-T^2}|\t_0\rangle +
\langle C\xi|\frac{f_e+T}{1-T^2}|\t_0\rangle \right.\0\\
&& \left. + (f_\e-1) \, \langle\xi|\frac T{1-T^2}|\t_+\rangle
- \langle\xi|\frac {f_e+T}{1-T^2}|\t_+\rangle \right]
|\xi+C\xi\rangle
\ee
The solution to this equation is clearly of the form $\t=\t_0+ H
|\xi+C\xi\rangle$, for some constant $H$. The latter can be determined by
plugging
this ansatz in (\ref{t+-'}). One easily gets
\beq
\t_+ \,=\, \t_0 + \frac 1{\k+f_e} |\xi+C\xi\rangle
\langle\xi|\frac 1{1+T}|\t_0\rangle\label{solt+}
\eeq
Now we can replace this solution back into (\ref{t++}). One easily
obtains
\beq
\t'_+ \,=\, \t_0 + \frac 1{\k+f_\e f_e}
|\xi+C\xi\rangle \langle\xi|\frac 1{1+T}|\t_0\rangle\label{solt+'}
\eeq
We see that as $\e\to 1$, $\t_+'\to\t_+$.

As for (\ref{t'-}) we proceed in the same way. From the difference equation
we obtain
\be \label{M-t-0} \!\!\!\!\!\!
M_- |\t_-\rangle \,\equiv\, \left[ 1 + \frac{1}{\k+f_\e f_e}\,
|\xi-C\xi\rangle \left((f_\e-1) \langle \xi| \frac T{1-T^2}
- \langle \xi| \frac {f_e+T}{1-T^2} \right) \right] |\t_-\rangle
\,=\, 0
\ee
The solution must be in the kernel of the operator $M_-$
and must have the form
\beq
|\t_-\rangle = \beta\, |(1-C)\xi\rangle \label{ansatzt-}
\eeq
for some constant $\beta$.
Plugging this in the previous equation we find
\be
M_- |\t_-\rangle \,=\, \b \frac{(f_\e-1)(f_e+\k)}{\k+f_\e f_e}
|\xi-C\xi\rangle \0
\ee
Therefore, (\ref{ansatzt-}) solves (\ref{M-t-0}) either when $f_\e=1$
($\e=1$), or when $f_e=-\k$ ($e\to\infty$) and $f_\e\ne1$. We are
interested here in the first case. Putting $f_\e=1$ and (\ref{ansatzt-})
in (\ref{t'-}) we obtain $\t'_-=\t_-$ for any $\b$ and $f_e$.

\newsection{Calculating $G$}

Let us first compute $G$ with $\t=\t_+$ starting from eq.(\ref{G1}).
Our procedure consists in separating the $\xi$--independent part
from the rest. The latter corresponds to Hata et al.'s
calculation, \cite{HKw, HM1, HM2}. For instance
\be
&&(\v_+-\v_-,\v_--\v_0) \,\hat\K_{\e e}^{-1}\,
\hat\T_{\e e}\, \left(
\matrix{\v_--\v_+\cr \v_+-\v_0\cr}\right)= (\v_+-\v_-,\v_--\v_0) \,
\K^{-1}\,\T\, \left(
\matrix{\v_--\v_+\cr \v_+-\v_0\cr}\right)\label{C1}\\
&& +(\v_+-\v_-,\v_--\v_0) \,\K^{-1}\,\frac 1{B_{\e e}}
\left(\matrix{e f_e &\e(\rho_1-\k\rho_2)\cr e(\rho_2-\k\rho_1)
&\e f_\e\cr}\right)
\CP_{\e e}(1+\M\K^{-1}\T)\left(
\matrix{\v_--\v_+\cr \v_+-\v_0\cr}\right)\0
\ee
where again $B_{\e e}= 1+(1-e)(1-\e)\k$. The first piece in the RHS is
the $\xi$--independent part. Carrying out the algebra one gets the
following result (\ref{C1})
\be
&&(\v_+-\v_-,\v_--\v_0) \,\hat\K_{\e e}^{-1}\,
\hat\T_{\e e}\, \left(
\matrix{\v_--\v_+\cr \v_+-\v_0\cr}\right)\,=\,
3\,\langle\t_0|\frac {T(2T-1)}{(T+1)^2(T-1)}|\v_0\rangle \label{C2}\\
&&~~~~~~~~~~~~~+\frac 2{B_{\e e}}
\left[ \langle\t_0|\frac 1{1-T^2}|\xi\rangle\left(
e(1-\e) \langle\xi|\frac T{1-T^2}|\t_0\rangle -
\e\langle\xi|\frac 1{1-T^2}|\t_0\rangle\right)\right]\0
\ee
Proceeding in the same way with the third term in (\ref{G1}) we find
\be
&&(\v_+-\v_-,\v_--\v_0)\,\hat\K_{\e e}^{-1}\,
\left(\matrix{0\cr \t_+}\right) =
\frac 12 \langle\t_0|\frac 1{1-T}|\t_0\rangle
+\frac 1{B_{\e e}}\left[e(\e-1) \langle\t_0|\frac T{1-T^2}|\xi\rangle
\langle\xi|\frac T{1-T^2}|\t_+\rangle\right.\0\\
&&~~~~~~~~+\left.\langle\t_0|\frac 1{1-T^2}|\xi\rangle
\left((\e-e(1-\e))\langle\xi|
\frac T{1-T^2}|\t_+\rangle +\e
\langle\xi|\frac 1{1-T^2}|\t_+\rangle\right)
\right]\label{C3}
\ee

Similarly for the last term on the RHS of (\ref{G1}) we find
\be
&&(0,\t_+) \,\M  \,\hat\K_{\e e}^{-1}\,
\left(\matrix{0\cr \t_+}\right) = \langle\t_0|\frac T{1-T^2}|\t_0\rangle+
\label{C4}\\
&&+ \frac 2{B_{\e e}}\left[e(1-\e)\langle\t_+|\frac T{1-T^2}|\xi\rangle
\langle\xi|\frac T{1-T^2}|\t_+\rangle
-\e \langle\t_+|\frac 1{1-T^2}|\xi\rangle
\langle\xi|\frac T{1-T^2}|\t_+\rangle\right]\0
\ee

Now we turn to the terms containing the twist--odd part. We need
\be
&& -2 (\v_+-\v_-,\v_--\v_0)\,\hat\K_{\e e}^{-1}\,
\left(\matrix{0\cr \t_-}\right)
-(0,\t_+) \,\M  \,\hat\K_{\e e}^{-1}\,
\left(\matrix{0\cr \t_-}\right)
+(0,\t_-) \,\M  \,\hat\K_{\e e}^{-1}\,
\left(\matrix{0\cr \t_+}\right) \0 \\
&& \qquad = -\frac{\beta(1-\e)}{1+(1-\e)(1-e)\kappa}
\left[(2+2\kappa -\e \k)\langle \t_0|\frac 1{1+T}|\xi\rangle
+e\kappa\langle \t_0|\frac 1{1-T}|\xi\rangle\right]
\label{C8}
\ee
and also
\beq
(0,\t_-) \,\M  \,\hat\K_{\e e}^{-1} \,
\left(\matrix{0\cr \t_-}\right) \,=\, 2\beta^2 \k
\frac{(1-\e)(\k+1)}{1+(1-\e)(1-e)\k}\label{C7}
\eeq

Using above formulae in (\ref{G1}) and (\ref{G2}) one obtains
(\ref{G1calc}) and (\ref{G2calc}), respectively.

\newsection{Formulas for star products in LEOM}

In this Appendix we explicitly write down some formulas which are
needed in order to evaluate the star products in the LEOM
when the involved state is of the type (\ref{ansatz}) with a nontrivial
polynomial $\EP$, or, in other words, is the product of a tachyon--like
state times a polynomial of the creation operators like (\ref{monom}).
The best course in this case is to introduce the state (\ref{genans}),
which depends on the variable vector $\b^\mu$, compute the star
products of this state with the dressed sliver and then differentiate
with respect to $\b^\mu$, setting $\b^\mu=0$ afterwards, in such a way as
to `pull down' the desired monomials of the type (\ref{monom}).
The calculation is straightforward and the relevant results for the matter
part are recorded
in the following formulas (where, for simplicity, we have set $\e=1$)
\be
&&(\ldots \langle \zeta a^\dagger_\mu\rangle\ldots)
\ket {\hat \varphi_e(\t,p)}*\ket{\hat \Xi} = \0
\\ &&\quad\quad =
(\ldots \langle -\zeta \frac{\partial}{\partial \b^\mu} \rangle
\ldots)
\,\exp\left[-\frac 12 \EA_1 - \EB_1 - p \cdot (\EC_1 + \ED_1)\right]
\ket {\hat \varphi_e(\t,p)} \, \Big\vert_{\b=0}
\label{star1}\\
&&\ket{\hat \Xi} * (\ldots \langle \zeta a^\dagger_\mu \rangle\ldots)
\ket {\hat \varphi_e(\t,p)}=
\0 \\
&&\quad\quad =
(\ldots \langle - \zeta \frac{\partial}{\partial \b^\mu}\rangle
\ldots)
\,\exp\left[-\frac 12 \EA_2 -\EB_2 - p \cdot (\EC_2 + \ED_2)\right]
\ket {\hat \varphi_e(\t,p)} \, \Big\vert_{\b=0}
\label{star2}
\ee
where
\be
&&\EA_1\equiv (\b,0) \,\M  \,\hat\K_{e 1}^{-1}\,
\left(\matrix{C\b\cr 0\cr}\right)\0\\
&&=\bra {\b} \frac T{1-T^2}\ket{C\b}- \bra {\b}  \frac 1{1-T^2}\ket {C\xi}
\bra {\xi} \frac T{1-T^2}\ket{\b}-\bra {\b}  \frac T{1-T^2}\ket {\xi}
\bra {\xi} \frac 1{1-T^2}\ket{C\b}\label{A1}\\
&&\EA_2\equiv (0,\b) \,\M  \,\hat\K_{1e}^{-1}\,
\left(\matrix{0\cr C\b\cr}\right)\0\\
&&=\bra {\b} \frac T{1-T^2}\ket{C\b}- \bra {\b}  \frac 1{1-T^2}\ket {\xi}
\bra {\xi} \frac T{1-T^2}\ket{C\b}-\bra {C\b}  \frac T{1-T^2}\ket {\xi}
\bra {\xi} \frac 1{1-T^2}\ket{\b}\label{A2}
\ee

\be
&&\EB_1\equiv a^\dagger(V_+,V_-) \,\hat\K_{e 1}^{-1}\,
\left(\matrix{C\b\cr 0\cr}\right)
= \, \langle a^\dagger \rho_2 \b \rangle + \frac 1{\k+f_e}
\langle a^\dagger C\xi\rangle \bra {\xi}\frac{T+f_e}{1-T^2}\ket{C\b}
\label{B1}\\
&&\EB_2\equiv a^\dagger(V_+,V_-) \,\hat\K_{1e}^{-1}\,
\left(\matrix{0\cr C\b\cr }\right)
=\, \langle a^\dagger \rho_1 \b\rangle + \frac 1{\k+f_e}
\langle a^\dagger \xi\rangle \bra {\xi}\frac{T+f_e}{1-T^2}\ket{\b}
\label{B2}
\ee

\be
&&\EC_1\equiv (\t,0) \,\M  \,\hat\K_{e 1}^{-1}\,
\left(\matrix{C\b\cr 0\cr}\right)\0\\
&&=\bra {\t} \frac T{1-T^2}\ket{C\b}- \bra {\t}  \frac 1{1-T^2}\ket {C\xi}
\bra {\xi} \frac T{1-T^2}\ket{\b}-\bra {\t}  \frac T{1-T^2}\ket {\xi}
\bra {\xi} \frac 1{1-T^2}\ket{C\b}\label{EC1}\\
&&\EC_2\equiv (0,\t) \,\M  \,\hat\K_{1e}^{-1}\,
\left(\matrix{0\cr C\b\cr}\right)\0\\
&&=\bra {\t} \frac T{1-T^2}\ket{C\b}- \bra {\t}  \frac 1{1-T^2}\ket {\xi}
\bra {\xi} \frac T{1-T^2}\ket{C\b}-\bra {C\t}  \frac T{1-T^2}\ket {\xi}
\bra {\xi} \frac 1{1-T^2}\ket{\b}\label{EC2}
\ee
and
\be
&&\ED_1\equiv (\v_+-\v_0,\v_--\v_+) \,\hat\K_{e 1}^{-1}\,
\left(\matrix{C\b\cr 0\cr}\right)\0\\
&&=- \bra{\t_0}\frac {\rho_1T+\rho_2}{1-T^2}\ket{C\b}+
\bra{\t_0}\frac 1{1-T^2}\ket{\xi}
\left[ \bra {\xi}\frac 1{1-T^2}\ket{C\b}+ \bra {\xi}  \frac T{1-T^2}
\ket {\b}\right]\label{D1}\\
 &&\ED_2\equiv (\v_+-\v_-,\v_--\v_0) \,\hat\K_{1e}^{-1}\,
\left(\matrix{0\cr C\b\cr}\right)\0\\
&&=- \bra{\t_0}\frac {\rho_1T+\rho_2}{1-T^2}\ket{\b}+
\bra{\t_0}\frac 1{1-T^2}\ket{\xi}
\left[ \bra {\xi}\frac 1{1-T^2}\ket{\b}+ \bra {\xi}  \frac T{1-T^2}
\ket {C\b}\right]\label{D2}
\ee

\newsection{Calculations for the vector state}

Applying the formulas of the previous section in the particular case
of the vector excitation (\ref{vectoransatz})
we get
\be
&&|\hat\varphi_{e,v} \rangle \,* \,|\hat \Xi\rangle +
|\hat \Xi\rangle \,*\,|\hat\varphi_{e,v} \rangle \,=\,
e^{-\frac 12 G\,p^2}
\left\{d^\mu \langle a^\dagger (1-C)\zeta\rangle\right.\0\\
&& + \left. \frac 1 {\k+f_e} d^\mu
\langle a_\mu^\dagger (1-C)\xi\rangle\,\bra{\xi}
\frac {f_e+T}{1-T^2}\ket{\zeta}
\,+\, p\cdot d\left[- \, 2\bra{\t}\frac T{1-T^2}\ket{(1-C)\zeta}
\right.\right.\label{E1}\\
&&+ \left.\left.
\langle\t|\frac T{1-T^2}|(1-C)\xi\rangle\bra {\xi}
\frac 1{1-T^2}\ket{\zeta}
+\langle\t|\frac 1{1-T^2}|(1-C)\xi\rangle\bra {\xi}
\frac T{1-T^2}\ket{\zeta}
\right]\right\} \N_v |\hat \varphi_e(\t,p)\rangle\0
\ee
A necessary condition to satisfy the LEOM is
\be
\bra{\xi} \frac {f_e+T}{1-T^2}\ket{\zeta}=0\0
\ee
On the other hand, the presence of the operator $1-C$ in all the terms
of the
second line tells us that only the $\t_-$ part of $\t$ contributes to this
terms. Inserting the explicit form of $\t_-$ one easily finds the
result (\ref{vectorsol}).

\newsection{Level 2 calculations}

Using the results of Appendix D, and keeping in mind the formulas
\be
&&\rho_1\ket{0_\pm}=\frac12\ket{0_\pm}+\frac12\ket{0_\mp}\0\\
&&\rho_2\ket{0_\pm}=\frac12\ket{0_\pm}-\frac12\ket{0_\mp}\0
\ee
the explicit formulas for the level 2 state are as follows
\be
&&\left( \theta^{\mu\nu} \langle a_{\mu}^\dagger \ket{\zeta_-}
\langle a_{\nu}^\dagger \ket{ \zeta_-}
\ket{\hat\varphi(\t,p)}\right)* \ket{\hat\Xi}+
\ket{\hat\Xi}*\left(\theta^{\mu\nu} \langle a_{\mu}^\dagger \ket{\zeta_-}
\langle a_{\nu}^\dagger \ket{ \zeta_-}
\ket{\hat\varphi(\t,p)}\right)\label{F1}\\
&&= e^{-\frac 12 Gp^2}\left[\frac 12\theta^{\mu\nu} \,
\langle a_{\mu}^\dagger \ket{\zeta_-}\,
\langle a_{\nu}^\dagger \ket{ \zeta_-}+ 2\, \theta_\mu^\mu \left(\,
\bra{\zeta_-} \frac T{1-T^2}
\ket{\zeta_-} + 2 \bra{\zeta_-} \frac 1{1-T^2} \ket{\xi}
\bra{\xi} \frac T{1-T^2}\ket{\zeta_-}\right)\right.\0\\
&& \quad + \,2\, \theta^{\mu\nu}\, \left(
\langle a_{\mu}^\dagger \ket{ \zeta_+} \,p_\nu\, \EH_+ +
\langle a_{\mu}^\dagger \ket{ \zeta_-} \,p_\nu\, \EH_- \right.\0\\
&&\quad\quad\quad\quad\left.\left. +\frac 1{\k+1}\langle
a_{\mu}^\dagger(1+C)\ket{ \xi}\,
\bra{\xi} \frac 1{1-T}\ket{\zeta_-}p_\nu \EH_+
+  p_\mu p_\nu (\EH_+^2 +\EH_-^2)\right)\right]\ket{\hat\varphi(\t,p)} \0
\ee
where we have used $\ket {\zeta_+}=(\rho_1-\rho_2)\ket{\zeta_-}$ and
we have disregarded terms that explicitly vanish when $\eta\to 0$,
i.e. evanescent terms like (\ref{vanishstate}). Moreover
\be
\EH_+&\!=\!& - \bra{\t_+}\frac T{1-T^2}\ket{\xi}
\bra{\xi}\frac 1{1-T^2}\ket{\zeta_-} + \bra{\t_+}\frac 1{1-T^2}\ket{\xi}
\bra{\xi}\frac T{1-T^2}\ket{\zeta_-}\label{H+}\\
&& + \frac 12
\bra{\t_0}\frac 1{1+T}\ket{\zeta_+} + \bra{\t_0}\frac 1{1-T^2}\ket{\xi}
\bra{\xi}\frac 1{1+T}\ket{\zeta_-}\0
\ee
and
\be
\EH_-= -\b \bra{\xi}\frac {T-\k}{1-T^2}\ket{\zeta_-}\label{H-}
\ee

The other relevant star product is
\be
&&\left(g^{\mu} \langle a_{\mu}^\dagger\ket{s_+}
\ket{\hat\varphi(\t,p)}\right) * \ket{\hat\Xi}+\ket{\hat\Xi} *
\left(g^{\mu} \langle a_{\mu}^\dagger\ket{s_+}  \ket{\hat\varphi(\t,p)}\right)
\label{F2}\\
&&=e^{-\frac 12 Gp^2}\left[g^{\mu} \langle a_{\mu}^\dagger\ket{s_+}
 + \frac 1{\k+1} g^\mu \langle a_{\mu}^\dagger(1+C)\ket{ \xi}
\bra{\xi}\frac 1{1-T}\ket{s_+}\right.\0\\
&& \quad - \, (p\cdot g) \left(\bra{\t_0} \frac 1{1-T} \ket {s_+}
- 2\, \bra{\t_0}\frac 1{1-T^2}\ket{\xi} \bra{\xi}\frac 1{1-T}\ket{s_+}
- 2\, \bra{\t_+} \frac{T}{1-T^2} \ket {s_+} \right.\0\\
&&\quad\quad\quad\left.\left. +
2\, \bra{\t_+}\frac 1{1-T^2}\ket{\xi} \bra{\xi}\frac T{1-T^2}\ket{s_+}
+ 2\, \bra{\t_+}\frac T{1-T^2}\ket{\xi}
\bra{\xi}\frac 1{1-T^2}\ket{s_+}\right)\right]
\ket{\hat\varphi(\t,p)} \0
\ee
In order for the LEOM to be satisfied the sum of (\ref{F1}) and
(\ref{F2}) must reproduce (\ref{level2st}). A first condition for this to be
true can be easily recognized: the coefficient in front of the
$\theta^{\mu\nu} \,\langle a_{\mu}^\dagger \ket{\zeta_-}\,
\langle a_{\nu}^\dagger \ket{ \zeta_-}$ term in the RHS of (\ref{F1})
must be 1, which implies $p^2= -1$. This identifies the mass
of the solution with the level 2 mass. Next, many terms in the
RHS of (\ref{F1},\ref{F2}) diverge as $\eta\to 0$. Therefore another condition
for LEOM to be satisfied is that the corresponding coefficients vanish.
Every bracket in the previous formulas are calculated by going to
the $k$--basis, i.e. by inserting a completeness $\int dk \ket{k}\bra {k}$
and then evaluating the $k$ integral.
The brackets that contain
$\ket{s_+},\ket{\eta_-},\ket{ \zeta_+}$ involve integrals evaluated
essentially at $k=0$; the other brackets are finite.
Remembering (\ref{expan'}), (\ref{as+}), (\ref{azeta-}) and moreover
that $\t_0$ is finite at $k=0$ (see Appendix A), while $\frac 1{1+T(k)} \sim 1/k$ and
$\xi(k)\to0$ as $k\to 0$ and $|k_0|>2\eta$, it is easy to determine
the degrees of divergence for $\eta\approx 0$.
To simplify the analysis we introduce an auxiliary assumption which
was already mentioned in the text.
We assume that $\xi(k)\neq 0$ only for $k<k_0<0$.
This makes all terms containing $\xi$ in the previous formulas irrelevant as far
as the LEOM is concerned. Under this hypothesis eq.(\ref{F1}) reduces to
(\ref{leom21}) and eq.(\ref{F2}) to (\ref{leom22}).
The surviving quantities are as follows
\be
&&\bra{\zeta_-} \frac T{1-T^2}\ket{\zeta_-}= -
\frac {\zeta_0^2\ln 3}{\pi}\frac 1{\eta^2}
-2 \frac {\zeta_0\zeta_1 \ln 3}{\pi}\frac 1{\eta} +
{\pi} \zeta_0^2 -
\frac {\ln 3}{\pi}(\zeta_1^2+2\zeta_0\zeta_2)+\ldots\label{z-z-}\\
&&\EH_+= \frac 12 \bra {\t_0} \frac 1{1+T} \ket{\zeta_+} + \ldots =
- \frac {\zeta_0 \ln 3}{\sqrt{\pi}}\frac 1{\eta} -
\frac {\zeta_1 \ln 3}{\sqrt{\pi}}+\ldots\label{H+app}\\
&&
\bra{\t_0} \frac 1{1+T}\ket{s_+}= -\frac{2 s_{-1} \ln 3}{\sqrt{\pi}}
\frac  1 {\eta^2}- \frac{2 s_{0} \ln 3}{\sqrt{\pi}}\frac  1 {\eta}-
(\frac 1{24} \sqrt{\pi^3}s_{-1} + \frac {2 s_1 \ln 3}{\sqrt {\pi}})
+\dots\label{s+t0}
\ee
It is important to notice that the numbers (in particular $\ln 3$)
that appear in this expansion depends heavily on the particular
regulator state $\ket{\eta}$ (\ref{epsi}) we are using. Therefore
they should not be attributed any particular significance. This
also imply that the conditions we will obtain below are
regularization dependent (see comment at the end section 7.1).

Now we can impose the necessary cancelations. We must have
\be
2\, \theta^{\mu\nu}\,
\langle a_{\mu}^\dagger \ket{ \zeta_+} \,p_\nu\, \EH_+
+\frac 12 g^{\mu} \langle a_{\mu}^\dagger\ket{s_+} =0\label{F3}
\ee
in the limit $\eta\to 0$. This implies that
$ g_\mu\sim \theta_{\mu\nu}p^\nu$. Assuming (\ref{Vir2'}) we find
\beq
s_{-1} = -2\sqrt{\frac2\pi} \, b \, \zeta_0^2 \ln 3 \;.
\label{1stcond}
\eeq

The next requirement is that
\beq
2\,\theta_\mu {}^\mu \bra{\zeta_-} \frac T{1-T^2}\ket{\zeta_-}+
2 \, \theta^{\mu\nu} p_\mu p_\nu \EH_+^2 -  p\cdot g \,
\bra {\t_0} \frac 1{1+T} \ket{s_+}=0
\label{F4}
\eeq
All three terms diverge like $\eta^{-2}$ as $\eta\to 0$.
The most divergent contribution
vanishes if $\,5ab \ln 3=4$. The vanishing of the $1/\eta$ term requires
\beq
\sqrt2\,\zeta_0\zeta_1\,(ab\ln 3-4) + \sqrt{\pi} a\,s_0 = 0
\label{F5}
\eeq
This equation binds together the values of $s_0,\zeta_0,\zeta_1$.
Finally we must impose that also the $\eta^0$ term vanishes.
This results in an equation of the same
type as (\ref{F5}), involving also $\zeta_2 $ and $s_1$. It is not very
illuminating and therefore we will not write it
explicitly.

After imposing these (mild) conditions we see that the linearized
EOM is satisfied provided $p^2=-1$ and the Virasoro constraints
in the form (\ref{Vir2'}) are satisfied.

To end this appendix, let us add a few lines on how one can
do without the auxiliary assumption made before eq.(\ref{z-z-}).
In this case we give up this assumption and simply take $\xi(k) \sim k$
as $k\to 0$ (this satisfies (\ref{noncond}) in a far less restrictive
way than the auxiliary condition). Then all the terms in the RHS of
(\ref{F1},\ref{F2}) are nonvanishing. Two types of terms are
dangerous: the term containing $\langle a^\dagger\ket{\zeta_-}$ in the
RHS of (\ref{F1}) and the two terms proportional to
$\langle a^\dagger (1+C)\ket{\xi}$, which are present in both equations.
These terms cannot be canceled within the present ansatz for the level
2 state. To deal with the first term we can add to the ansatz
(\ref{level2st})
a term $g^{\mu} \langle a_{\mu}^\dagger\ket{r_-}\ket{\hat\varphi(\t,p)}$
where $\ket{r_-}$ is similar to $\ket{\zeta_-}$, and
$r(\eta)= r_0+r_1\eta+\ldots$. Adjusting the parameter $r_0$ we can easily
cancel the first dangerous term. As for the other two, we can simply
add to the ansatz two terms formally equal to the two terms of
(\ref{level2st}), where $\ket{\zeta_-}$ and $\ket{s_+}$ are replaced by
$\ket{(1-C)\zeta'}$ and $\ket{(1+C)s'}$, with $\rho_2\zeta'=\zeta',
\rho_1\zeta'=0$ and $\rho_2 r'=r',\rho_1 r'=0$. We can easily take
$\zeta'(k),r'(k)$ to cancel the above two terms as well as
all the remaining terms not containing string oscillators $a^\dagger$.

\newsection{Level 3 calculations}

The first part of this appendix is devoted to redefining the polarizations
as mentioned at the beginning of section 7.2. Such redefinitions are
as follows
\be
&&h_\mu = A\, g_\mu+B\, p\cdot g \,p_\mu\0\\
&& \l_{\mu\nu} = C\,\omega_{\mu\nu} +(D_+\,p_\mu  \omega_{\rho\nu}+
D_-\,p_\nu \omega_{\mu\rho})p^\rho  +D' \,p_\mu p_\nu \omega^{\rho\sigma}p_\rho
p_\sigma\label{AG1}\\
&& \chi_{\mu\nu\rho} = E\,\theta_{\mu\nu\rho}+ F\,(p_\mu
\theta_{\sigma\nu\rho} + p_\nu  \theta_{\mu\sigma\rho} +
  p_\rho\theta_{\mu\nu\sigma}) p^\sigma \0\\
&& \quad\quad\quad +H\,( p_\mu p_\nu \theta_{\sigma\tau\rho}+
p_\mu p_\rho \theta_{\sigma\nu\tau}+
p_\nu p_\rho \theta_{\mu\sigma\tau})p^\sigma p^\tau
+ H'\, p_\mu p_\nu p_\rho \theta^{\l\sigma\tau}p_\l p_\sigma p_\tau\0
\ee

Inserting the above redefinitions into (\ref{Vir3'a},\ref{Vir3'b})
we get
\be\label{AG2}
&& 3\sqrt{2}\,\left(\frac {A-2B}{C} - 2\,\frac AC \,\frac {D_++D_--D'}{D_+-2D'}
\right)
g\cdot p +\,2\,\omega_\mu{}^\mu =0\0\\
&&3\,g_\mu +\sqrt{2} \frac{C-2D_-}{A} \omega_\mu{}^\nu p_\nu=0
\label{Vir3'}\\
&& 2\sqrt{2}\frac CE\,\omega_{\nu\mu} p^\nu
 - \sqrt{2}\left( \frac {C+4D_+-2D_-}E +\frac {(D_++D_-)(F-H)}{E(F-2H)} \right)
\omega_{\mu\nu} p^\nu +3 \,\theta_{\mu\nu}{}^\nu =0\0\\
&& 2\, \omega_{(\mu\nu)} +3\sqrt{2} \,\frac {E-2F}{C}\, \theta_{\mu\nu\rho}p^\rho=0\0
\ee
These equations are of the same form as (\ref{Vir3a},\ref{Vir3b})
with an obvious identification of the coefficients $x,y,u,v,z$. The
coefficients $A,\ldots,H'$ are subject to the conditions
\be
&&\frac {E-2F}{C} =\frac {H-2H'}{D'},\quad\quad \frac {E-2F}{2C} =
\frac{F-2H}{D_++D_-},\quad\quad \frac{C-2D_-}{A} =\frac{D_+-2D'}{B}\0\\
&&F\left(C+2D_--4D_+ - \frac{(F-H)(D_++D_-)}{F-2H}\right)
= E\left( 2D_--D_+ -2D'-\frac{2D'(H-H')}{H-2H'}\right)\0
\ee

The second part of the Appendix concerns the equations that must
be verified among the terms of eqs.(\ref{thetastar3},\ref{thetalambdastar3})
and (\ref{starr3}) for the LEOM (\ref{verlevel3}) to be satisfied.
As explained in the text we have to impose that all the terms
in the RHS of eqs.(\ref{thetastar3},\ref{thetalambdastar3})
and (\ref{starr3}) that do not reproduce the level 3 state vanish.
There are two such terms: one linear in $a^\dagger$
\be
&&3\, \theta_\mu{}^{\mu\rho} \,\bra{\zeta_-}
\frac T{1-T^2}\ket{\zeta_-}\,\langle a_\rho^\dagger\ket{\zeta_-}
+3\,\theta^{\mu\nu\rho} \langle a_{\rho}^\dagger
\ket{\zeta_-}\,p_\mu\,p_\nu\, \EH_+^2 \0\\
&&+ \omega^{\mu\nu} \,\left(\langle a_{\mu}^\dagger \ket{ \zeta'_-} p_\nu\,
\bra{\t_0}\frac T{1-T^2}
\ket{\l_+}\, + \,\langle a_{\nu}^\dagger \ket{ \l_-} p_\mu\,\EH_+\right)
+\frac 34 \,g^\mu\, \langle a_\mu^\dagger \ket{r_-}=0\label{G3}
\ee
and another quadratic in $a^\dagger$
\be
3\,\theta^{\mu\nu\rho}\, \langle a_{\mu}^\dagger \ket{\zeta_-}
\langle a_{\nu}^\dagger \ket{\zeta_+} p_\rho \EH_+
+ \omega^{\mu\nu} \left(\frac 14\,
\langle a_{\mu}^\dagger \ket{\zeta'_-}\,
\langle a_{\nu}^\dagger \ket{ \l_+}+ \frac 12\,
\langle a_{\mu}^\dagger \ket{\zeta'_+}\,
\langle a_{\nu}^\dagger \ket{ \l_-}\right)=0\label{G4}
\ee
Now we use the $\eta$--expansions (\ref{z-z-}) and (\ref{H+app}), together
with
\be
\bra{\t_0} \frac T{1-T^2}\ket{\l_+}= \frac{ \l_{-1} \ln 3}{\sqrt{\pi}}
\frac  1 {\eta^2}+ \left(\frac{\l_{0} \ln 3}{\sqrt{\pi}}-
\frac {\sqrt{\pi}}4\l_{-1}\right)\frac  1 {\eta}-
\left(\frac   {\sqrt{\pi^3}}{48}\l_{-1} - \frac {\l_1 \ln 3}{\sqrt {\pi}}
+\frac {\sqrt{\pi}}4 \l_0\right)
+\dots\0
\ee
Equation (\ref{G4}) implies $2\,\omega_{(\mu\nu)} +
3\sqrt{2}\, z\,\theta_{\mu\nu}{}^\rho p_\rho=0$ for some $z$.
The terms in the RHS of (\ref{G4}) are of overall order 0 in $\eta$,
therefore only one condition is requested:
\beq
3\,\sqrt{2}\,z\,\zeta_0'\, \l_{-1}= 8\,\frac{\zeta_0^3}{\sqrt{\pi}} \ln 3\label{Gfirst}
\eeq
The RHS of (\ref{G3}) contains
terms of order --2,--1 and 0 in $\eta$ as $\eta\to 0$. We must therefore
satisfy three conditions. Using that $\theta^{\mu\nu\rho}p_\mu p_\nu\sim
\omega^{(\rho\mu)}p_\mu$, we see that the condition involving the term
of order --2, takes exactly the form of the first equation (\ref{Vir3b})
with
\be
&&u= \sqrt{\frac {\pi}{2}} \,\frac {\zeta_0'\l_{-1}}{\zeta_0^3} - \frac {\ln 3}
{6z}\label{u}\\
&& v= \frac 12\, \sqrt{\frac {\pi}{2}} \, \frac {\l_1}{\zeta_0^2} + \frac {\ln 3}
{12z}\label{v}
\ee
For generic values of $\zeta_0,\zeta_0',\l_{-1}$, eqs.(\ref{Gfirst},\ref{u},\ref{v})
fix $u,v$ and $z$ to some specific (nonvanishing) values.
Now the vanishing of the term $\sim\eta^{-1}$ leads to an equation
similar to the first equation (\ref{Vir3b}), with identifications
for $u$ and $v$ different from (\ref{u},\ref{v}),
\be
&&u=-\frac 23\, \sqrt{\frac {\pi}{2}} \,
\frac {\zeta_0'\l_0 +\zeta_1'\l_{-1}}{\zeta_0^2\zeta_1} - \frac {\ln 3}
{6z}\label{u'}\\
&& v= \frac 1{6\sqrt{2}}\,   \frac {\l_1\zeta_1 +\l_0\zeta_0}
{\zeta_1 \zeta_0^2} + \frac {\ln 3} {12z}\label{v'}
\ee
These equations however involve three additional parameters
$\zeta_1,\l_0,\zeta_1'$.
So it is easy to tune them to the same specific values for $u$ and $v$.
Finally the term of order $\eta^0$ involves also $g^\mu$. In this case
there are several possibilities\footnote{In order to restrict
the number of these possibilities and obtain more binding
conditions one should give up the simplifying
assumption and treat the level 3 in full generality.}.
One of these is that $g_\mu \sim
\omega_{\nu\mu}p^\nu$. In the latter case also the constant $y$ in
(\ref{Vir3a}) gets determined in terms of all the parameters, which include
now also $\zeta_2,\zeta_2',\l_1,r_0$. Since the relevant equations are
cumbersome and not particularly illuminating we do not write them down.
In conclusion the LEOM for
the state (\ref{level3st}) is satisfied together with the Virasoro
constraints (\ref{Vir3a},\ref{Vir3b}) (the first is a consequence of the
other three), provided some mild conditions on the various parameters
that enter the game be complied with.

\newsection{The cochain space}
In this Appendix we would like to explain in more detail the definition
of the space of cochains given in section 8.

From eq.(\ref{coho1}), it would seem that,
should we keep $\eta$ finite throughout the cohomological analysis,
all the states we have constructed would be trivial. This is due
to the fact that in the gauge transformed expressions there appears
the operator $\rho_1-\rho_2$, which has the property that
$(\rho_1-\rho_2)\ket{\eta_{\pm}} =\ket{\eta_\mp}$. However this would be
misleading, since in this argument we forget all the corrections to
the LEOM that vanish only when $\eta\to 0$.
In addition one should not forget that the $\eta$ dependence is an artifact
of our regularization, it does not correspond to anything that has
to do with the
physical string modes. It can only appear at an intermediate step in our
calculations. Therefore the space of cochains should not contain any
reference to
the $\eta$ dependence. There are only two ways to implement this. We can say that
every cochain is defined up to evanescent states. But this would lead to
incurable inconsistencies: for instance, the 0 state would be defined up to
evanescent states, but we know that by applying, for instance,
a gauge transformation to
$\ket{\eta_+}$, which is evanescent, we get $\ket{\eta_-}$,
which is in a nonzero class; so we would get the paradoxical result that
applying the BRST operator to 0, we get something different from zero.
This possibility has consequently to be excluded.

The only consistent possibility is the one put forward in the text.
The nonzero cochains are those obtained by explicitly taking the limit $\eta\to 0$
for non-evanescent states, that is taking the limit in
expressions of the type $\langle a^\dagger \zeta\rangle$
both for regular and singular $\zeta$'s (see section 6, in particular
formulas (\ref{nepsi})). It is clear that one gets in this way  well--defined
expressions for the states. This will form the set of nonzero cochains.
To this we have to add the zero cochain, which is simply the zero state.
All together they form a linear space. By definition this is the space
of cochains where we want to compute the cohomology of the VSFT fluctuations.
The $\eta$--regularization
enters into the game when we come to compute the star products of the LEOM or of
(\ref{gt}). Without such regularization these star products are ambiguous.
From this point of view we see that the $\eta$--regularization concerns
the star product rather than the states themselves. The cohomological
problem at this point is well--defined.

\newsection{Towers of solutions}

In this appendix we prove the statements used in section 8 to show that
for any solutions to the LEOM we can construct an infinite tower
of solutions with the same mass. We begin with the calculation
of the star product $(h^\nu\langle a^\dagger_\nu
\xi\rangle\ket{\hat\varphi(\theta,n,\t,p)})* \ket{\hat \Xi}$.
Written down explicitly this becomes
\be
&&\left(h^\nu\langle a^\dagger_\nu \xi\rangle\sum_{i=1}^n
\theta^{\mu_1\ldots\mu_i}_i \langle a_{\mu_1}^\dagger \zeta^{(i)}_1\rangle
\ldots \langle a_{\mu_i}^\dagger \zeta^{(i)}_i\rangle
\ket{\hat\varphi(\t,p)}\right) * \ket{\hat \Xi}
\,=\, h^\nu \sum_{i=1}^n \theta^{\mu_1\ldots\mu_i}_i \times
\label{T1} \\
&& \times \langle -\xi \frac {\partial}{\partial \b^\nu}\rangle
\langle -\zeta^{(i)}_1 \frac {\partial}{\partial \b^{\mu_1}}\rangle
\ldots
\langle -\zeta^{(i)}_i \frac {\partial}{\partial \b^{\mu_i}}\rangle
\, \exp\!\left[-\frac 12 \EA_1-\EB_1 - p\,(\EC_1+\ED_1) \right]
\ket{\hat\varphi(\t,p)} \, \Big\vert_{{\bf\beta}=0} \0
\ee
Now we set $ \EF_1 = -\frac 12 \EA_1$ and
$\EG_1 =-\EB_1 - p\,(\EC_1+\ED_1)$. Then the RHS of (\ref{T1}) becomes
\be
&&= h^\nu \sum_{i=1}^n \theta^{\mu_1\ldots\mu_i}_i
\langle -\zeta^{(i)}_1 \frac {\partial}{\partial \b^{\mu_1}}\rangle
\ldots
\langle -\zeta^{(i)}_i \frac {\partial}{\partial \b^{\mu_i}}\rangle
\langle \xi \frac{\partial (\EF_1+\EG_1)}{\partial \b^\nu}\rangle
\, {\rm e}^{[\EF_1+\EG_1]} \Big\vert_{\b=0} \, \ket{\hat\varphi(\t,p)}
\0 \\
&& = h^\nu \sum_{i=1}^n \theta^{\mu_1\ldots\mu_i}_i
\langle -\zeta^{(i)}_1 \frac {\partial}{\partial \b^{\mu_1}}\rangle
\ldots
\langle -\zeta^{(i)}_i \frac {\partial}{\partial \b^{\mu_i}}\rangle
\langle -\xi \frac {\partial \EG_1}{\partial \b^\nu}\rangle
\, {\rm e}^{[\EF_1+\EG_1]} \Big\vert_{\b=0} \, \ket{\hat\varphi(\t,p)} +
\0 \\
&& +\, h^\nu \sum_{i=1}^n \theta^{\mu_1\ldots\mu_i}_i \,
\sum_{j=1}^i \left(\langle \xi \frac {\partial^2 \EF_1}
{\partial\b^\nu\,\partial\b^{\mu_j}}\zeta^{(i)}_j\rangle
\langle -\zeta^{(i)}_1 \frac {\partial}{\partial \b^{\mu_1}}\rangle
\ldots \widetilde{\langle -\zeta^{(i)}_j
\frac{\partial}{\partial \b^{\mu_j}}\rangle} \ldots
\langle -\zeta^{(i)}_i \frac {\partial}{\partial \b^{\mu_i}}\rangle
\right) \times
\0 \\
&& \qquad \times \, {\rm e}^{[\EF_1+\EG_1]} \Big\vert_{\b=0} \,
\ket{\hat\varphi(\t,p)}
\0 \\
&&= h^\nu \langle -\xi \frac {\partial \EG_1}{\partial \b_\nu}\rangle
\left( \sum_{i=1}^n
\theta^{\mu_1\ldots\mu_i}_i \langle a_{\mu_1}^\dagger \zeta^{(i)}_1\rangle
\ldots \langle a_{\mu_i}^\dagger \zeta^{(i)}_i\rangle \ket{\hat\varphi(\t,p)}
*\ket{\hat \Xi}\right)\0\\
&& +h^\nu \sum_{j=1}^i \langle \xi \frac {\partial^2 \EF_1}
{\partial\b^\nu\,\partial\b^{\mu_j}}\zeta^{(i)}_j\rangle
\left(\theta^{\mu_1\ldots\mu_j\ldots\mu_i}_i
\langle a_{\mu_1}^\dagger \zeta^{(i)}_1\rangle\ldots
\widetilde{\langle a_{\mu_j}^\dagger \zeta^{(i)}_j\rangle}
\ldots \langle a_{\mu_i}^\dagger \zeta^{(i)}_i\rangle \ket{\hat\varphi(\t,p)}
*\ket{\hat \Xi}\right) \0
\ee
Tilded quantities denote omitted ones.
Now, using the formulas of Appendix D, and the fact that
$\rho_2\xi=\xi, \rho_1\xi =0$, it is easy to prove that
\be
\langle -\xi \frac {\partial \EG_1}{\partial \b_\nu}\rangle
=\langle a^\dagger_\nu \xi\rangle,\quad\quad
\langle \xi \frac {\partial^2 \EF_1}
{\partial\b^\nu\,\partial\b^{\mu_j}}\zeta^{(i)}_j\rangle =
\eta_{\nu\mu_j} \bra{\xi} \frac {\k-T}{1-T^2}
\ket {C\zeta^{(i)}_j}\0
\ee
Inserting this back in the previous equations, we obtain
\be
&&\left(h^\nu\langle a^\dagger_\nu \xi\rangle\sum_{i=1}^n
\theta^{\mu_1\ldots\mu_i}_i \langle a_{\mu_1}^\dagger \zeta^{(i)}_1\rangle
\ldots \langle a_{\mu_i}^\dagger \zeta^{(i)}_i\rangle
\ket{\hat\varphi(\t,p)}\right) * \ket{\hat \Xi} =
\0 \\
&&=\, h^\nu\langle a^\dagger_\nu \xi\rangle \left[\sum_{i=1}^n
\theta^{\mu_1\ldots\mu_i}_i \langle a_{\mu_1}^\dagger \zeta^{(i)}_1\rangle
\ldots \langle a_{\mu_i}^\dagger \zeta^{(i)}_i\rangle \ket{\hat\varphi(\t,p)}
*\ket{\hat \Xi}\right]\label{T2}\\
&& +\, h^\nu \sum_{j=1}^i \eta_{\nu\mu_j} \bra{\xi} \frac {\k-T}{1-T^2}
\ket {C\zeta^{(i)}_j}
\left(\theta^{\mu_1\ldots\mu_j\ldots\mu_i}_i
\langle a_{\mu_1}^\dagger \zeta^{(i)}_1\rangle\ldots
\widetilde{\langle a_{\mu_j}^\dagger \zeta^{(i)}_j\rangle}
\ldots \langle a_{\mu_i}^\dagger \zeta^{(i)}_i\rangle \ket{\hat\varphi(\t,p)}
*\ket{\hat \Xi}\right) \0
\ee

Now we repeat the calculation for the commuted product
\be
&&\ket{\hat \Xi}*\left(h^\nu\langle a^\dagger_\nu \xi\rangle\sum_{i=1}^n
\theta^{\mu_1\ldots\mu_i}_i \langle a_{\mu_1}^\dagger \zeta^{(i)}_1\rangle
\ldots \langle a_{\mu_i}^\dagger \zeta^{(i)}_i\rangle
\ket{\hat\varphi(\t,p)}\right)
\,=\, h^\nu \sum_{i=1}^n \theta^{\mu_1\ldots\mu_i}_i \times
\label{T3} \\
&& \; \times\, \langle -\xi \frac {\partial}{\partial \b^\nu}\rangle
\langle -\zeta^{(i)}_1 \frac {\partial}{\partial \b^{\mu_1}}\rangle
\ldots
\langle -\zeta^{(i)}_i \frac {\partial}{\partial \b^{\mu_i}}\rangle
\, \exp\!\left[-\frac 12 \EA_2-\EB_2 - p\,(\EC_2+\ED_2)\right]
\ket{\hat\varphi(\t,p)} \, \Big\vert_{{\bf\beta}=0} \0
\ee
Now, to simplify notation, we set $\EF_2 = -\frac 12 \EA_2$ and
$\EG_2 =-\EB_2 -p\,(\EC_2+\ED_2)$, and proceeding as before (\ref{T3})
becomes
\be
&&= h^\nu \langle -\xi \frac {\partial \EG_2}{\partial \b_\nu}\rangle
\left( \ket{\hat \Xi}* \left(\sum_{i=1}^n
\theta^{\mu_1\ldots\mu_i}_i \langle a_{\mu_1}^\dagger \zeta^{(i)}_1\rangle
\ldots \langle a_{\mu_i}^\dagger \zeta^{(i)}_i\rangle \ket{\hat\varphi(\t,p)}\right)
\right)\0\\
&& +h^\nu \sum_{j=1}^i \langle \xi \frac {\partial^2 \EF_2}
{\partial\b^\nu\,\partial\b^{\mu_j}}\zeta^{(i)}_j\rangle
\left(\ket{\hat \Xi}*\left(\theta^{\mu_1\ldots\mu_j\ldots\mu_i}_i
\langle a_{\mu_1}^\dagger \zeta^{(i)}_1\rangle\ldots
\widetilde{\langle a_{\mu_j}^\dagger \zeta^{(i)}_j\rangle}
\ldots \langle a_{\mu_i}^\dagger \zeta^{(i)}_i\rangle \ket{\hat\varphi(\t,p)}\right)
\right) \0\\
&&=h^\nu\langle a^\dagger_\nu \xi\rangle \left[\ket{\hat \Xi}*\left(\sum_{i=1}^n
\theta^{\mu_1\ldots\mu_i}_i \langle a_{\mu_1}^\dagger \zeta^{(i)}_1\rangle
\ldots \langle a_{\mu_i}^\dagger \zeta^{(i)}_i\rangle \ket{\hat\varphi(\t,p)}\right)
\right]\label{T4}
\ee
since
\be
\langle -\xi \frac {\partial \EG_2}{\partial \b_\nu}\rangle
=\langle a^\dagger_\nu \xi\rangle,\quad\quad
\langle \xi \frac {\partial^2 \EF_2}
{\partial\b^\nu\,\partial\b^{\mu_j}}\zeta^{(i)}_j\rangle=0\0
\ee

Collecting the above results we have
\be
&&\left(h^\nu\langle a^\dagger_\nu
\xi\rangle\ket{\hat\varphi(\theta,n,\t,p)}\right)* \ket{\hat \Xi}+
\ket{\hat \Xi}*\left(h^\nu\langle a^\dagger_\nu
\xi\rangle\ket{\hat\varphi(\theta,n,\t,p)}\right) =
\0 \\
&& =\, h^\nu\langle a^\dagger_\nu\xi\rangle\left[
\ket{\hat\varphi(\theta,n,\t,p)}* \ket{\hat \Xi}+
\ket{\hat \Xi}*\ket{\hat\varphi(\theta,n,\t,p)} \right] +
\label{T5} \\
&& +\, h^\nu \sum_{j=1}^i \eta_{\nu\mu_j} \bra{\xi} \frac {\k-T}{1-T^2}
\ket {C\zeta^{(i)}_j}
\left(\theta^{\mu_1\ldots\mu_j\ldots\mu_i}_i
\langle a_{\mu_1}^\dagger \zeta^{(i)}_1\rangle\ldots
\widetilde{\langle a_{\mu_j}^\dagger \zeta^{(i)}_j\rangle}
\ldots \langle a_{\mu_i}^\dagger \zeta^{(i)}_i\rangle \ket{\hat\varphi(\t,p)}
*\ket{\hat \Xi}\right) \0
\ee
The last line vanishes if
$\bra{\xi} \frac{T-\k}{1-T^2} \ket{C\zeta^{(i)}_j}=0$ or if, for those
$j$'s for which this is not true, $h$ is transverse to the tensor
$\theta$. In this case, if $\ket{\hat\varphi(\theta,n,\t,p)}$ is a
solution to the linearized equation of motion,
\be
\left(h^\nu\langle a^\dagger_\nu
\xi\rangle\ket{\hat\varphi(\theta,n,\t,p)}\right)* \ket{\hat \Xi}+
\ket{\hat \Xi}*\left(h^\nu\langle a^\dagger_\nu
\xi\rangle\ket{\hat\varphi(\theta,n,\t,p)}\right) =
h^\nu\langle a^\dagger_\nu
\xi\rangle\ket{\hat\varphi(\theta,n,\t,p)}\label{T6}
\ee
i.e. also $h^\nu\langle a^\dagger_\nu
\xi\rangle\ket{\hat\varphi(\theta,n,\t,p)}$ is a solution. All the results
similar to this used in section 8 can be obtained by obvious extensions
of the previous argument.

\newsection{Calculating $H$}
\label{app:Hcalc}

The number $H$ comes from the three--point tachyon vertex. If we take
(\ref{etachyon}) as the tachyon solution, the three--tachyon vertex is
given by
\beq
{}_1\langle\phi_e(\t,p_1)|_2\langle\phi_e(\t,p_2)
|_3 \langle\phi_e(\t,p_3)| V_3\rangle_{123} \,=\,
\left(\det \,\hat \K_3\right)^{-\frac D2}
\hat \N_e^3\, \exp [-{\cal H}_1(p_1,p_2,p_3)] \label{3t}
\eeq
${\cal H}_1$ is given by
\beq
{\cal H}_1(p_1,p_2,p_3)= \chi^T\hat\K_3^{-1} \lambda +\frac 12 \lambda^T
\V_3\hat\K_3^{-1}\lambda + \frac 12\chi^T\hat\K_3^{-1}\hat\Sigma_3\chi +\frac 12
(p_1^2+p_2^2+p_3^2)V_{00} \label{H1}
\eeq
with $p_1+p_2+p_3=0$. In this equation the various symbols are as follows
\be
&& \lambda^T=(\l_1,\l_2,\l_3), \quad\quad \l_i =-p_i \t C,\quad i=1,2,3\0\\
&&\chi = \left(\matrix{\v_0p_1+\v_-p_2+\v_+p_3\cr \v_+p_1+\v_0p_2+\v_-p_3\cr
\v_-p_1+\v_+p_2+v_0p_3\cr}\right)= \left(\matrix{\v_0&\v_-&\v_+\cr
\v_+&\v_0&\v_-\cr\v_-&\v_+&\v_0\cr}\right)
\left(\matrix{p_1\cr p_2\cr p_3}\right)\label{notat}\\
&&\hat\Sigma_3=\left(\matrix{\hat S_e &0&0\cr 0&\hat S_e &0\cr
0&0&\hat S_e \cr}\right),\quad\quad \V_3 =\left(\matrix{V&V_+&V_-\cr
V_-&V&V_+\cr V_+&V_-&V\cr}\right)\0
\ee
Finally $\hat{\K}_3 = 1-\hat \Sigma_3 \V_3$.
Since
\beq
\N_e \,=\, \frac{g_0}{\sqrt{G}} \sqrt{\frac{\det(1-\hat S_e^2)^{
\frac D2}}{\det(1-\widehat {\tilde S}^2)}} \exp[- \frac 12 p^2\t
\frac 1{1+\hat T_e}C\t]
\label{Nt}
\eeq
the total exponential in (\ref{3t}) is given by
\beq
{\cal H} = {\cal H}_1+{\cal H}_2, \quad\quad {\cal H}_2(p_1,p_2,p_3)=
(p_1^2+p_2^2+p_3^2) H_2,\quad\quad H_2 =-\frac 12
\langle\t| \frac 1{1+\hat T_e}C|\t\rangle\label{H2}
\eeq
Similarly one can show that
${\cal H}_1 (p_1,p_2,p_3) = (p_1^2+p_2^2+p_3^2)
H_1$. Let us set $H=H_1+H_2$.

All expressions can be straightforwardly computed once we explicitly
determine the quantity
\be
\hat\K^{-1}_3=\left(1-\hat\Sigma_3 \V\right)^{-1}\0
\ee
It turns out that
\be
\hat\K^{-1}_3=\K_3^{-1}\left[1+\left(1-\EP_e\M_3\K_3^{-1}
\right)^{-1}\EP_e\M_3\K_3^{-1}
\right]
\ee
Moreover we have
\be
&&\left(1-\EP_e\M_3\K_3^{-1}\right)^{-1}\EP_e=\sum_{n=0}^\infty
\left(\EP_e\M_3\K_3^{-1}\right)^n\EP_e\0\\
&&=\frac{\k+f_e}{f_e^3-1}\left(\matrix{f_e^2 & (f_e\rho_1+\rho_2)
& (f_e\rho_2+\rho_1)\cr
(f_e\rho_2+\rho_1) & f_e^2 & (f_e\rho_1+\rho_2)\cr
(f_e\rho_1+\rho_2) & (f_e\rho_2+\rho_1) & f_e^2 \cr}\right)\EP_e
\ee
With the use of this formula one can directly compute all the
contributions in (\ref{H1}) given the general tachyon solution
\be
\t&=&\t_++\t_-\0\\
\t_+&=&\t_0+ \a\,W \left(\xi +C\xi\right),
\quad W=\langle\xi|\frac 1{1+T}|\t_0\rangle\\
\t_-&=& \beta \left(\xi -C\xi\right)\0
\ee
When momentum conservation holds we have the following identity
\be
(p_1,p_2,p_3)\left(\matrix{a & b & c \cr c & a & b \cr b & c & a}\right)
\left(\matrix{p_1\cr p_2\cr p_3}\right){\Bigg |}_{\sum_i p_i=0}=
\left(a-\frac12(b+c)\right)\sum_i p_i^2
\ee

Let us begin analyzing the contribution coming from the twist--even part of
tachyon. With lenghty but straightforward manipulations we get
\be
\chi^T\hat\K_3^{-1} \lambda&=&-\frac 12(p_1^2+p_2^2+p_3^2)
{\Big[}\bra{\t_0}\frac{1}{1-T^2}\ket{\t_+}\\
&&-\frac{2}{f_e^2+f_e+1} \bra{\t_0}\frac{1}{1-T^2}\ket{\xi}
\left((f_e-1)\bra{\xi}\frac{T}{1-T^2}\ket{\t_+}-(1+2f_e)\bra{\xi}
\frac{1}{1-T^2}\ket{\t_+}\right){\Big]}\0\\
\0\\
\lambda^T\V_3\hat\K_3^{-1}\lambda &=&(p_1^2+p_2^2+p_3^2){\Big[}\bra{\t_+}
\frac{T}{1-T^2}\ket{\t_+}-\frac12
\bra{\t_+}\frac{1}{1-T^2}\ket{\t_+}\\
&&+\frac{1}{f_e^2+f_e+1}{\Big(}(f_e-1)\bra{\t_+}\frac{1}{1-T^2}\ket{\xi}^2-
(2f_e+1)\bra{\t_+}\frac{T}{1-T^2}\ket{\xi}^2\0\\
&&+2(f_e+2)\bra{\t_+}\frac{1}{1-T^2}\ket{\xi}\bra{\t_+}\frac{T}{1-T^2}
\ket{\xi}{\Big)}{\Big]}\0\\
\0\\
\chi^T\hat\K_3^{-1}\hat\Sigma_3\chi&=&\frac 32 (p_1^2+p_2^2+p_3^2)
{\Big(}\bra{\t_0}\frac{T(1-2T)}{(1-T^2)(1+T)}\ket{\v_0}
+\frac{f_e+2}{f_e^2+f_e+1}\bra{\t_0}\frac{1}{1-T^2}\ket{\xi}^2{\Big)}
\ee
Plugging inside the expression for $\t_+$ we get
\be
{\cal H}_1^+(p_1,p_2,p_3)&=&(p_1^2+p_2^2+p_3^2){\Big\{}\frac 12\,\bra{\t_0}
\frac{1}{1+T}\ket{\t_0}+2\a W^2+\a^2(\k-\frac 12)W^2\\
&&-\frac{1}{f_e^2+f_e+1}\,\frac 12{\Big[}\a^2\left(f_e(1+2\k-2\k^2)
-(1-4\k+\k^2)\right)\0\\
&&+2\a \left(f_e(2\k-1)+(\k-2)\right)-(2f_e+1){\Big]}W^2{\Big\}}\0
\ee
The second contribution to $H$ comes from the normalization in front of the
 tachyon state (\ref{Nt}), that is
\be
{\cal H}_2^+(p_1,p_2,p_3)=
-\frac 12\,(p_1^2+p_2^2+p_3^2)\,\langle\t_+| \frac 1{1+\hat T_e}C|\t_+\rangle\0
\ee
We have
\be
&&\langle\t_+| \frac 1{1+\hat T_e}C|\t_+\rangle=\bra{\t_+}\frac{1}{1+T}
\ket{\t_+}+\frac{2}{\k+f_e}\bra{\t_+}\frac{1}{1+T}\ket{\xi}^2
\sum_{n=0}^{\infty} \left(\frac{\k-1}{\k+f_e}\right)^n\\
&&=\bra{\t_0}\frac{1}{1+T}\ket{\t_0}+4\a W^2+2\a^2(\k-1)W^2
+\frac{2}{f_e+1}\left(\a(\k-1)+1\right)^2 W^2
\ee
The total twist--even contribution in $H$, let us call it $H^+$, is then
\be
{H}^{+} &=& {H}_1^{+}+{ H}_2^{+}\\
&=& H_0 - \frac{(f_e-1)^2(\k+f_e)^2}{2(f_e+1)(f_e^3-1)}
\left(\frac{1}{\k+f_e}-\a\right)^2
\langle\t_0|\frac 1{1+T}|\xi\rangle^2
\ee
 The bare contribution $H_0$ is naively zero but, in level truncation
regularization, it
acquires a non--vanishing value, \cite{HM2}. We stress once more that
the dressing contribution
is not affected by twist anomaly as the half string vector $\xi$
does not excite
the $k=0$ (zero momentum) midpoint mode.

Now we turn to the twist--odd contributions which, for $e\neq1$, does
not vanish identically for any solution to the LEOM. Let's analyze first
the purely
imaginary contribution linear in $\b$. It is easy to see that the part
coming from
$H_2$ is identically zero by twist symmetry, and the same is true for
the term
$\lambda^T_-\V_3\hat\K_3^{-1}\lambda_+$ in $H_1$. So the only potential
contributions arise from the tachyon linear term
$\chi^T\hat\K_3^{-1} \lambda_-$.
It is straightforward to compute these terms  by plugging $\t_-= \beta
\left(\xi -C\xi\right)$ and  to show again that twist symmetry requires this
contribution to vanish. So there are no imaginary contributions in $H$.\\
The  quadratic terms in $\b$ come out from $\lambda^T_-\V_3\hat\K_3^{-1}
\lambda_{-}$ in $H_1$ and
$\langle\t_-| \frac 1{1+\hat T_e}C|\t_-\rangle$ in $H_2$. They can be
directly computed plugging the explicit expression for $\t_-$. The result is
\be
\lambda^T_-\V_3\hat\K_3^{-1}\lambda_-&=&\b^2\frac{(f_e+\k)(f_e(2\k-1)+
(\k-2))}{f_e^2+f_e+1}\sum_i p_i^2\\
-\frac 12\langle\t_-| \frac 1{1+\hat T_e}C|\t_-\rangle &=&
-\beta^2\frac{(\k-1)(\k+f_e)}{f_e+1}
\ee
Together they sum up to
\be
{\cal H}^{-}&\equiv& H^-\, \sum_i p_i^2= \frac12 \lambda^T_-\V_3\hat\K_3^{-1}\lambda_-
- \frac12\langle\t_- | \frac 1{1+\hat T_e}C|\t_-\rangle \sum_i p_i^2 \\
H^-&=& \b^2 \frac{(f_e-1)^2(\k+f_e)^2}{2(f_e+1)(f_e^3-1)}
\ee

\newsection{Role of the critical dimension}

This Appendix is devoted to the role played by the critical dimension
(D=26) in VSFT, see section 10. Let us start from the normalized action
\beq\label{Sbmp}
S[\hat\psi]=-\frac{1}{g_0^2}\left(\frac{1}{2}\bra{\hat\psi}{\cal
Q} \ket{\hat\psi} +
\frac13\bra{\hat\psi}\hat\psi*\hat\psi\rangle\right) \eeq By means
of the operator field redefinition \cite{Oka2} \beq\label{redef}
\psi=e^{-\frac14\ln\gamma(K_2-4)}\hat\psi \eeq it can be brought
to the form \beq\label{Shk2}
S'[\psi]=-\frac{1}{g_0^2\gamma^3}\left(\frac12\bra\psi{\cal
Q}\ket{\psi} + \frac13\bra\psi\psi*\psi\rangle\right)=
\,-\frac{1}{g_0^2}\left(\frac{1} {2\gamma}\bra{\tilde\psi}{\cal
Q}\ket{\tilde\psi}+\frac13\bra{\tilde\psi}
\tilde\psi*\tilde\psi\rangle\right) \eeq where $\tilde \psi =
\gamma \psi$. Both forms of the action have been considered
previously in the literature,  \cite{GRSZ2, HKw}, in the limit
$\gamma\to 0$, implying a singular normalization of the action.
What we have shown above is that free effective parameters appear
in the process of regularizing the classical action so that a
singular normalization of the latter can be avoided. This remark
is of more consequence than it looks at first sight. The point is
that the redefinition (\ref{redef}) can harmlessly be implemented
only in $D=26$. In noncritical dimensions, as a consequence of
such a redefinition, an anomaly appears, \cite{GJ1}. In the course
of our derivation above the critical dimension has never featured,
but this remark brings it back into the game. This has an
important consequence: setting $\gamma=g_0^{2/3}$ in the middle
term of eq.(\ref{Shk2}), it is evident that in critical dimensions
we can make any parameter to completely disappear from the action
by means of a field redefinition. So, in $D=26$, the value of the
brane tension is dynamically produced and not put in by hand.

The very reason for this is that the family of operators
$K_n=L_n-(-1)^nL_{-n}$ leaves the action cubic
term invariant (only in D=26) while it acts linearly on the kinetic
term as, \cite{Oka2}
\beq
[K_{2n},{\cal Q}]=-4n(-1)^n{\cal Q}\label{KQ}
\eeq
In other words $\Q$ is an ``eigenvector'' of $K_{2n}$, and so every
parameter can be absorbed by a field redefinition.
In OSFT, on the other hand,  one cannot implement a redefinition like
(\ref{redef}) since $Q_B$ does not transform as an eigenvector of
$K_{2n}$, so the coupling constant there is really a free parameter
in the action.

Let us elaborate more on this aspect. We remark that both the
string fields $\psi$ and $\hat \psi$ above satisfy the same EOM. Therefore
there seems to exist different solutions of the EOM corresponding to
the same energy, and, on the other hand, a given solution can be
attributed different tensions
(depending on what constant we put in front of the action, which does not
affect the EOM). Since any constant put in front of the action in VSFT
in critical dimension can be absorbed via a field redefinition, it is
illusory to try to cure this problem by multiplying the action by some
constant. This is a fact of VSFT in critical
dimension and we have to come to terms with it. It is
apparent from the above that VSFT does not predict the exact value of
the D-brane tension, but rather makes room for it to emerge dynamically.
It is at this point that dressing comes handy.
We showed in [1] that in the theory there naturally arise scaling constants
$s$ and $\tilde s$ (see eq. (6.25) there) that can be adjusted to
the physical value of the D-brane tension. Therefore the answer to the
above puzzle is that if we redefine the string field in the action
as in eq.(2.38),
the parameters $s$ and $\tilde s$ should be scaled accordingly in
such a way as to preserve the physical value of the brane tension.
Of course, in this way, we are left with a multiplicity
of solutions corresponding to the same tension. This has to be attributed
to an invariance of the type discussed in section 9 and 11 below
(perhaps a remnant of the original gauge invariance of the theory).

\newpage


\end{document}